\documentclass[journal,10pt,draftclsnofoot,onecolumn]{IEEEtran}
\usepackage{graphicx}
\usepackage{amssymb}

\ifCLASSOPTIONcompsoc
\else
\fi
\ifCLASSINFOpdf
\else
\fi
\input{tcilatex}
\begin{document}

\title{Structural Diversity for Resisting Community Identification in
Published Social Networks}
\author{Chih-Hua Tai, Philip S. Yu,~\IEEEmembership{Fellow,~IEEE,} \\De-Nian
Yang,~\IEEEmembership{Senior Member,~IEEE} and Ming-Syan Chen,~%
\IEEEmembership{Fellow,~IEEE}%
\IEEEcompsocitemizethanks{
\IEEEcompsocthanksitem C.-H. Tai is with
the Department of Computer Science and Information Engineering,
National Taipei University, New Taipei 23741, Taiwan. E-mail:
hanatai@mail.ntpu.edu.tw.
\IEEEcompsocthanksitem P. S. Yu is with the Department of Computer Science, University of Illinois at Chicago, IL 60607, USA. E-mail: psyu@cs.uic.edu.
\IEEEcompsocthanksitem D.-N. Yang is with
the Institute of Information Science and the Research Center of
Information Technology Innovation, Academia Sinica, Taipei 11529, Taiwan. E-mail: dnyang@iis.sinica.edu.tw.
\IEEEcompsocthanksitem M.-S. Chen is with the Research Center of
Information Technology
Innovation, Academia Sinica, Taipei 11529, Taiwan, and the Department of Electrical Engineering, National Taiwan University, Taipei 10617, Taiwan. E-mail: mschen@citi.sinica.edu.tw.}%
\thanks{}}

\IEEEcompsoctitleabstractindextext{\begin{abstract}

As an increasing number of social networking data is published and
shared for commercial and research purposes, privacy issues about
the individuals in social networks have become serious concerns.
Vertex identification, which identifies a particular user from a
network based on background knowledge such as vertex degree, is one
of the most important problems that has been addressed. In reality,
however, each individual in a social network is inclined to be
associated with not only a vertex identity but also a community
identity, which can represent the personal privacy information
sensitive to the public, such as political party affiliation. This
paper first addresses the new privacy issue, referred to as
community identification, by showing that the community identity of
a victim can still be inferred even though the social network is
protected by existing anonymity schemes. For this problem, we then
propose the concept of \textit{structural diversity} to provide the
anonymity of the community identities. The $k$-Structural Diversity
Anonymization ($k$-SDA) is to ensure sufficient vertices with the
same vertex degree in at least $k$ communities in a social network.
We propose an Integer Programming formulation to find optimal
solutions to $k$-SDA and also devise scalable heuristics to solve
large-scale instances of $k$-SDA from different perspectives. The
performance studies on real data sets from various perspectives
demonstrate the practical utility of the proposed privacy scheme and
our anonymization approaches.

\end{abstract}

}
\maketitle


\IEEEdisplaynotcompsoctitleabstractindextext

%
\IEEEpeerreviewmaketitle

\section{Introduction}

\label{1} \baselineskip=0.216in

%
%

%
%
%
%

\vspace{-0.5mm}
\begin{figure}[h]
\centering
\includegraphics[width=0.2\textwidth]{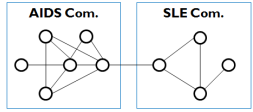} 
\caption{Privacy violation by degree attacks.}
\label{intro}
\end{figure}

\IEEEPARstart{I}{n} a social network, individuals are represented by
vertices, and the social activities between individuals are summarized by
edges. In light of the recognition of the usefulness of information in
social networking data for commercial and research purposes, more and more
social networking data have been published and shared in recent years. This,
however, raises serious privacy concerns for the individuals whose personal
information is contained in social networking data.

Each individual in a social network is associated with a vertex identity,
which can represent the user name or Social Security number (SSN)\footnote{%
%
SSN is a nine-digit number issued to U.S. citizens, permanent and temporary
residents in the United States.}. Vertex identification, where malicious
attackers utilize their background knowledge to associate an individual with
a specific vertex in published social networking data, is one of the most
important privacy issues that has emerged in recent year \cite{survey2,
survey1}. Due to the complexity of social networks, the resistance of vertex
identification has been studied against different background knowledge from
various perspectives \cite{backstorm, grouping, k-deg, k-neighbor, k-automor}%
. Backstrom et al. in \cite{backstorm} first showed that as long as an
attacker knows a piece of information about an individual, it is
insufficient to protect privacy by only removing the vertex identities. Liu
and Terzi in \cite{k-deg} later proposed $k$-degree anonymity that
guarantees the privacy protection against degree information. Given the
degree information, $k$-degree anonymity ensures that there are at least $k$
vertices with the same degree in a social network, such that the probability
of an individual being associated with a specific vertex is limited to $1/k$%
. 
Similar concepts have also been applied to provide protection against
attackers with stronger background knowledge. The work in \cite{k-neighbor}
considered the case where an attacker's knowledge is the 1-neighborhood
connectivity around an individual and proposed $k$-neighborhood anonymity as
a solution. The studies in \cite{k-iso, k-automor} introduced $k$%
-automorphism anonymity and $k$-isomorphism anonymity against attacks of
arbitrary subgraphs related to an individual. Alternatively, a
generalization technique is another approach. Hay et al. \cite{grouping}
were able to hide privacy details about each individual by grouping a set of
vertices into a super-vertex and inferring the relationships between
super-vertices from super-edges.

Note that, however, each individual in a social network is inclined to be
associated with a community identity \cite{x2, x1}. The community identity
of a vertex can represent the personal privacy information sensitive to the
public, such as on-line political activity group, on-line disease support
group information, or friend group association in a social network.
Different from the other vertex features such as gender or salary, community
identity is a kind of structural information that can be derived by the
community detection techniques from a social network. The existing vertex
anonymity schemes thus cannot ensure the privacy protection for the
community identities since it is possible that the vertices with the same
information known to an attacker gather closely in a subgraph (community) of
the whole social network.

Specifically, this paper addresses a new privacy issue, referred to as
community identification, and shows that $k$-degree anonymity is not
sufficient. Consider the $2$-degree anonymity in Figure \ref{intro} as an
example. Suppose that an attacker knows that John has 5 friends in this
network. In the case of explicit communities, the attacker is able to infer
that John has AIDS since all vertices with degree 5 are associated with the
AIDS community. Moreover, even in another case of implicit communities
(i.e., without explicit community label), the attacker can infer the
neighborhood of John with only a distance one inaccuracy by identifying the
dense subgraph in which John resides.
This example demonstrates that 
even though an attacker cannot precisely identify the vertex corresponding
to an individual, private and sensitive community information and
neighborhood information can still be revealed.



To prevent community identification in published social networks by degree
attacks, therefore, we propose $k$-structural diversity, which ensures that
for each vertex, there are other vertices with the same degree located in at
least $k-1$ different communities. The rationale is that the probability for
an attacker to associate a victim with the correct community identity is
limited to at most $1/k$. We then formulate a new problem, $k$-Structural
Diversity Anonymization ($k$-SDA), which ensures the $k$-structural
diversity with minimal semantic distortion. For $k$-SDA, we propose an
Integer Programming formulation to find optimal solutions for small
instances. In addition, we also devise scalable heuristics to solve
large-scale instances of $k$-SDA with different perspectives.
To demonstrate the practical utility of the proposed privacy scheme
and our anonymization approaches, various evaluations are performed
on real data sets. The experimental results show that the social
networks anonymized by our approaches can preserve much of the
characteristics of the original networks.




\vspace{-2mm}

\section{Related Work}


Privacy is always a crucial factor in releasing or exchanging data. In the
past decade, issues on privacy-preserving data publishing (PPDP) on
transaction data, such as record linkage, sensitive attribute linkage, and
table linkage, 
have attracted extensive research interest \cite{db-survey}. Record
linkage refers to the identification of a record's owner, and its
corresponding privacy model, $k$-anonymity \cite{k-anon1}, prevents
record linkage by ensuring that at least $k$ records share the same
quasi-identifier. That is, there are at least $k$ records in a
\textit{qid} group. Following this initial research, a group of
studies, such as MultiRelational $k$-anonymity \cite{multi-k},
extended $k$-anonymity to improve and support privacy protection
under various scenarios and attacks. In contrast to the record, the
attribute value associated with each individual is more important in
sensitive attribute linkage, and $l$-diversity \cite{l-div} ensures
that at least $l$ sensitive values appear in every \textit{qid}
group. However, as Li et al. \cite{t-close} observed, $l$-diversity
is not sufficient to provide privacy protection, especially when the
overall distribution of the sensitive attribute is skewed. In other
words, an attacker is able to issue a \textit{skewness attack} when
a sensitive attribute is associated to a \textit{qid} group with
higher confidence than other \textit{qid} groups. This problem is
remedied by $t$-Closeness \cite{t-close} by demanding that the
distribution of a sensitive attribute in every \textit{qid} group is
similar to each other among the whole dataset. It is worth noting that both $%
l$-diversity and $t$-closeness mainly focus on categorical sensitive
attributes. For numerical sensitive attributes, a \textit{proximity attack}
\cite{proxi} identifies the interval in which the sensitive value $s$ of an
individual is located, while $(\varepsilon ,m)$-anonymity is proposed to
ensure that the probability to infer an interval $[s-\varepsilon
,s+\varepsilon ]$ is limited to at most $1/m$. Moreover, table linkage is
concerned about whether the record associated with an individual is
presented in a released table, and $\delta$-presence \cite{presence} limits
the probability of the above inference within a specified range.

With the explosive growth of information from social networking
applications, privacy concerns in releasing social networking data become
increasingly important. Various issues, such as vertex identification and
link identification, have drawn extensive research interests \cite{survey2,
survey1}. Vertex identification \cite{backstorm, k-iso, grouping, k-deg,
k-neighbor, k-automor} finds the one-to-one correspondence of each
individual and each vertex in a social network in order to extract sensitive
personal information, and many anonymization and generalization approaches
for resisting vertex identification have been introduced in Section I. This
contrasts with link identification \cite{k-iso, RandSP, EdgeAnon, SensiShip}%
, which discloses the sensitive relationship between two individuals. To
resolve this issue, perturbation \cite{RandSP} with edge addition, edge
deletion, and edge swap is proposed. To further address different privacy
requirements, edges are classified into multiple types of sensitivities and
removed with different priorities \cite{SensiShip}. Zhang et al. \cite%
{EdgeAnon} explored a new situation where attackers possess the knowledge of
vertex descriptions, such as degrees, and proposed to decrease the certainty
on the existence of an edge according to the attacker's available knowledge.
In addition, $\alpha $-proximity \cite{tclose-in-sn} brings the notion of
attribute privacy in transaction data to social networks by extending the
concept of $t $-closeness. That is, $\alpha $-proximity ensures that the
distribution of labels in a neighborhood is similar to that in the whole
social graph.

Different from all the above privacy models concentrating on varied datasets
that are directly made public, \textit{differential privacy} \cite{dp}
explores the condition on the release mechanism, i.e., a randomized
algorithm $\mathcal{A}$ answering queries to release information.
Specifically, a randomized algorithm $\mathcal{A}$ follows $\epsilon $%
-differential privacy if for all datasets $x$ and $x^{\prime }$ that differ
on at most one element, and any subset of outputs $S\subset Range(\mathcal{A}%
)$,%
\[
Pr[\mathcal{A}(x)\in S]\leq exp(\epsilon )Pr[\mathcal{A}(x^{\prime })\in S],
\]
\noindent where $\epsilon $ is a privacy parameter. Intuitively, the
privacy protection increases with a smaller $\epsilon $. Thus,
differential privacy aims to introduce noises into query results and
provide robust privacy guarantee without any assumption on the data
and background knowledge possessed by an attacker. In the past few
years, the great promise of differential privacy has mainly been
demonstrated on statistical database \cite{db-survey}. Very
recently, a few studies \cite{dp-cut, dp-dd, dp-triangle} have also
proposed its application to social networks. To meet the privacy
guarantee, those approaches focus on specific data utility of social
networks. Specifically, Hay et al. \cite{dp-dd} proposed constrained
inferences to provide provable privacy for the degree distribution
of a social network; Karwa et al. \cite{dp-triangle} studied the
privacy-preserving problem for subgraph counting queries, e.g., a
triangle, k-star and k-triangle, while Gupta et al. \cite{dp-cut}
addressed the cut function of a graph that answers the number of
correspondences between any two sets of individuals.

\vspace{-2mm}

\section{Problem Formulation}

In this paper, we formulate a new anonymous problem, $k$-Structural
Diversity Anonymization ($k$-SDA), to protect the community identities of
individuals in a network. The network is represented as an undirected simple
graph $G(V,E,C)$, where $V$ is the set of vertices corresponding to the
individuals, $E$ is the set of edges representing the relationship between
individuals, and $C$ is the set of communities. These communities can be
either explicitly given as input or derived through clustering on the social
network graph. Each vertex $v$ has a community ID\footnote{%
For simplicity, we focus on the one-community case in this paper while the
multi-community scenario is studied in our ICDM paper \cite{ICDM11}.}, $%
c_{v} $, in $C$, and each edge in $E$ can span two vertices in either the
same or different communities. Let $d_{v}$ denote the degree of vertex $v $,
and $k$-SDA is also given a positive integer parameter $k$, $1\leq k\leq $ $%
|C|$, to represent the structural diversity, which is formally defined as
follows.\smallskip

\textbf{Definition 1.} A graph $G(V,E,C)$ is $k$-structurally diverse, i.e.,
satisfying $k$-SDA, if for every vertex $v\in V$, there exist at least $k$
communities such that each of the communities contains at least one vertex
with the degree identical to $d_{v}$\footnote{
From the viewpoint of privacy protection, the concept of structural
diversity proposed in this paper can be extended to support the
multi-community scenario \cite{ICDM11}. For protecting a
\textit{single} community, the structural diversity anonymization
($k$-SDA) specifies that the vertices of the same degree need to
appear in at least $k$ different communities. In contrast, to
support the scenario that each individual belongs to a community set
with one or more than one community, the key factor for extending
$k$-SDA is to ensure that the vertices of the same degree in the
anonymized graph appear in at least $k$ different mutually exclusive
\textit{community sets}. For example, if two vertices A and B of the
same degree in the anonymized graph belong to community sets \{{C1\}
and \{C2},{C3\},} respectively, those two vertices follow 2-SDA
since the two community sets are mutually exclusive. On the other
hand, if A and B reside in community sets \{{C1}\} and \{{C1,C3\},}
respectively, it is easy for an attacker to infer that A and B must
participate in community C1.}.\smallskip

\begin{figure}[tbh]
\begin{center}
$%
\begin{array}{ccc}
\includegraphics[width=0.07\textwidth]{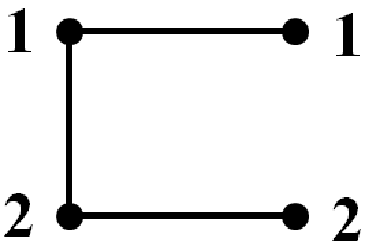} &  & %
\includegraphics[width=0.07\textwidth]{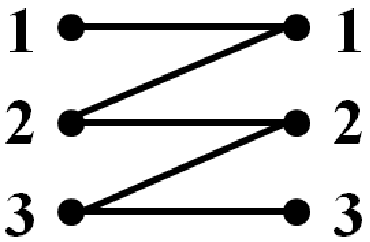} \\
{(a)} &  & {(b)}%
\end{array}
$%
\end{center}
\par
\vspace{-2mm}
\caption{Examples of two 2-structurally diverse graphs.}
\label{ex}
\end{figure}

\begin{figure}[h]
\centering
\includegraphics[width=0.155\textwidth]{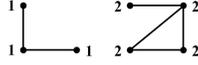}
\vspace{-1mm}
\caption{Examples of limit of operation Adding Edge.}
\label{unavailable}
\end{figure}

\begin{figure}[!tbh]
\begin{center}
$%
\begin{array}{ccccc}
\includegraphics[width=0.075\textwidth]{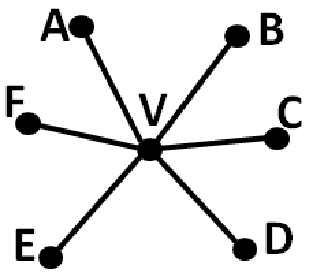} & %
\includegraphics[width=0.075\textwidth]{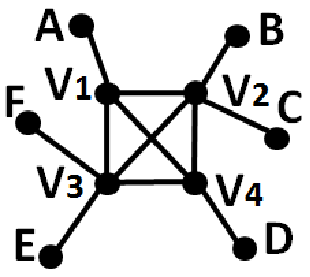} & %
\includegraphics[width=0.075\textwidth]{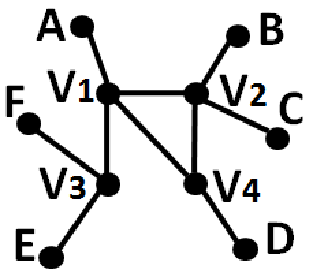} & %
\includegraphics[width=0.075\textwidth]{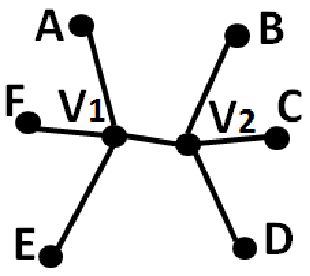} & %
\includegraphics[width=0.075\textwidth]{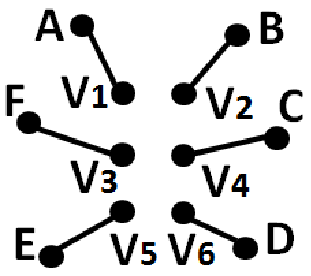} \\
{(a)} & {(b)} & {(c)} & {(d)} & {(e)}%
\end{array}
$%
\end{center}
\par
\vspace{-2mm} \caption{Examples of operation Splitting Vertex}
\label{splitting-ex}
\end{figure}

In other words, for each vertex $v$, there must exist at least $k-1$ other
vertices located in at least $k-1$ other communities. Figure \ref{ex} shows
an example with the graphs that are 2-structurally diverse, where the
community ID is indicated beside each vertex. In Figure \ref{ex}(a), both
communities contain a vertex with the degree as 1 and a vertex with the
degree as 2. Therefore, the graph is 2-structurally diverse. In Figure \ref%
{ex}(b), two communities contain vertices with the degree as 1, and three
communities contain vertices with the degree as 2. For each degree, we can
find at least two communities containing vertices with the same degree. The
graph is thus 2-structurally diverse.\smallskip

\textbf{Proposition 1.} If $G(V,E,C)$ is $k$-structurally diverse, then it
also satisfies $k$-degree anonymity, which implies that for every vertex,
there exist at least $k-1$ other vertices with the same degree.\smallskip

\textbf{Proposition 2.} If $G(V,E,C)$ is $k_{1}$-structurally diverse, then
it is also $k_{2}$-structurally diverse for every $k_{2}$, $k_{2}\leq k_{1}$%
.\smallskip

The problem is to anonymize a graph $G(V,E,C)$ such that the graph is $k$%
-structurally diverse. To limit the semantic distortion in the corresponding
applications, we define two operations, \textit{Adding Edge} and \textit{%
Splitting Vertex}. Operation Adding Edge connects two vertices belonging to
the same community. Adding an edge for two vertices in different communities
is prohibited because it may lead to improper distortion. 
For example, it is inappropriate to artificially connect an individual in
the liberal political action community to another individual in the
anti-abortion community to achieve $k$-structural diversity. Although
operation Adding Edge alone can fulfill $k$-structural diversity in some
cases, $k$-structural diversity cannot often be solely achieved with this
operation. Consider the example in Figure \ref{unavailable}. There is one
vertex with the degree as 3 in community 2. However, by operation Adding
Edge alone, it is impossible to make any vertex in community 1 have a degree
as 3 since there are only three vertices in community 1.

Therefore, operation Splitting Vertex is proposed to ensure that any
arbitrary input instance can be anonymized to achieve $k$-structural
diversity. Each vertex $v$ involved in this operation is split into
multiple \textit{substitute vertices}, where each substitute vertex
is a clone for the corresponding individual. Each clone represents
the relationship of at least one neighbor of $v$, such that all
substitute vertices of $v$ as a whole share the same relationships
with the neighbors of $v$ before the splitting. Specifically, let
$E_{v}$ denote the set of incident edges of $v$, where $v$ is
replaced with a set $S_{v}$ of substitute vertices such that (1)
each substitute vertex is connected with at least one edge in
$E_{v}$, and (2) every edge in $E_{v}$ is incident to a substitute
vertex in $S_{v}$. Thus, $S_{v}$ includes at most $\left\vert
E_{v}\right\vert $ vertices. Figures \ref{splitting-ex}(b)-\ref{splitting-ex}%
(e) present several possible results for Splitting Vertex on vertex
$v$ of Figure \ref{splitting-ex}(a). For the connectivity between
substitute vertices, a simple approach is to enforce that all
substitute vertices of $v$ must be mutually connected. However,
Splitting Vertex does not restrict that $S_{v}$ must form a clique
because an attacker can regard the clique as a hint to identify the
corresponding individual. Therefore, Splitting Vertex allows a
substitute vertex to freely connect to any other substitute vertex
in $S_{v}$, and the flexibility inherited in Splitting Vertex
enables our algorithm to achieve $k$-structural diversity for any
arbitrary input instance.

Note that in the previous study on the privacy preservation of databases
\cite{db-survey}, it was pointed out that maintaining the original
information stored in the database is important for some applications that
are required to extract the attribute values associated with the data
tuples. For this reason, several database anonymization schemes \cite%
{db-survey, t-close, l-div, k-anon2} avoid removing a tuple or even any of
its attribute values in order to preserve all corresponding information.
Similarly, for preserving the attribute values of a tuple to some extent,
many existing anonymization schemes \cite{t-close, l-div, k-anon2} adopt
generalization or suppression to hide a specific attribute value into its
specific attribute range or generalize the concepts of the attribute values,
while the hiding ranges and generalization concepts are optimized to reduce
the distortion.

In this paper, the proposed algorithms with operations Adding Edge and
Splitting Vertex can be regarded as the above type of anonymization schemes
that aims to preserve the attribute values to some extent. As such, the
information in the social networks is not removed by deleting or swapping
the existing edges, even though the above two strategies allow the proposed
algorithms to be more flexible in anonymizing a graph. Nevertheless, the
concept of swapping an edge has been incorporated in our algorithm design.
The proposed heuristics redirect an edge added at the previous iteration,
instead of always adding a new edge, in order to reduce the number of
created edges. However, redirecting added edges does not affect the original
edges in the network, and hence does not violate our objective of preserving
the original edges in the network. Specifically, the objective of $k$-SDA is
to minimize the semantic distortion during the anonymization via Adding Edge
and Splitting Vertex. We formally define $k$-SDA as follows.\smallskip

\textbf{Problem }$k$\textbf{-SDA.} Given a graph $G(V,E,C)$ and an integer $%
k $, $1\leq k\leq $ $|C|$, the problem is to anonymize $G$ to satisfy $k$%
-structural diversity with operations Adding Edge and Splitting Vertex such
that $n_{a}+\omega n_{s}$ is minimized, where $n_{a}$ denotes the number of
edges created in operation Adding Edge, $n_{s}$ denotes the number of
vertices added in operation Splitting Vertex, and $\omega $ is a positive
weight for operation Splitting Vertex.\smallskip

In this paper, we set $\omega $ as $\left\vert V\right\vert ^{2}$ (the
maximum number of edges in a graph) to consider the case that operation
Splitting Vertex is performed only if the graph cannot be anonymized with
operation Adding Edge alone.

\section{Integer Programming}

In the following, we propose the Integer Programming formulation for $k$%
-SDA. Our formulation together with any commercial software for mathematical
programming can find the optimal solutions, which can be used as the
benchmarks for the solutions obtained by any heuristic algorithm. We first
derive the formulation for $k$-SDA with only operation Adding Edge in
Section \ref{IP-1} to capture the intrinsic characteristics of this
optimization problem and to avoid initially including complicated details.
Thereafter, we extend the formulation to incorporate both operations in \ref%
{IP-2}.

\subsection{Formulation with Adding Edge}

\label{IP-1} 

\begin{table}[tbp] \centering%
\caption{The input of $k$-SDA.}%
\begin{tabular}{l|l}
\hline\hline
Notation & Description \\ \hline
$V$ & the set of vertices \\
$C$ & the set of communities \\
$E$ & the set of the original edges \\
$E_{v}$ & the set of the original edges incident on $v$ \\
& $v\in V$, $E_{v}\subseteq E$ \\
$\overline{E}$ & the set of candidate edges that are allowed to \\
& be added in operation Adding Edge \\
$\overline{E}_{v}$ & the set of adding edge candidates incident on $v$, \\
& $v\in V$, $\overline{E}_{v}\subseteq \overline{E}$ \\
$S_{v}$ & the set of substitute vertices of $v$, $v\in V$ \\
$D$ & the set of degrees, i.e., $D=\left\{ d\in
\mathbb{N}
\right. \left\vert 1\leq d\leq \left\vert V\right\vert \right\} $ \\
$k$ & the size of structural diversity \\
$c_{u}$ & the community of vertex $u$, $u\in V$, $c_{u}\in C$ \\ \hline\hline
\end{tabular}%
\label{Notation}%
\end{table}%

\begin{table}[tbp] \centering%
\caption{The decision variables of $k$-SDA with operation Adding Edge.}%
\begin{tabular}{l|l}
\hline\hline
Notation & Description \\ \hline
$\alpha _{u,v}$ & binary variable; $\alpha _{u,v}=1$ if edge $e_{u,v}$ is
added in \\
& operation Adding Edge; otherwise, $\alpha _{u,v}=0$, \\
& $e_{u,v}\in \overline{E}_{u}$ \\
$\delta _{u,d}$ & binary variable; $\delta _{u,d}=1$ if the degree of $u$ is
$d$; \\
& otherwise, $\delta _{u,d}=0$, $u\in V$, $d\in D$ \\
$\theta _{c,d}$ & binary variable; $\theta _{c,d}=1$ if there exists at least
\\
& one vertex in $c$ with its degree as $d$; otherwise \\
& $\theta _{c,d}=0$, $c\in C$, $d\in D$ \\ \hline\hline
\end{tabular}%
\label{Variable}%
\end{table}%
%
%

As an initial basis, consider the formulation for $k$-SDA with only
operation Adding Edge. Tables \ref{Notation} and \ref{Variable} summarize
the input and decision variables of $k$-SDA. In our formulation, $e_{u,v}$
and $e_{v,u}$ correspond to the same edge. The objective function of $k$-SDA
with only operation Adding Edge is formulated as%
\[
\min \sum\limits_{e_{u,v}\in \overline{E}}\alpha _{u,v}.
\]%
The objective function minimizes the number of added edges. The problem has
the following constraints,

\begin{eqnarray}  \label{(1)}
\forall u\in V, & &
\quad\quad\quad\quad\quad\quad\quad\quad\quad\quad\quad\quad\quad\quad
\end{eqnarray}
\[
\sum\limits_{d\in D}\delta _{u,d}=1,
\]
\begin{eqnarray}  \label{(2)}
\forall u\in V,\forall d\in D,\quad\ \quad\quad\quad\quad\
\quad\quad\quad\quad\quad& & \\
\mbox{ where }d<\left\vert E_{u}\right\vert \mbox{ or }d>\left\vert
E_{u}\right\vert +\left\vert \overline{E}_{u}\right\vert, \quad\ \ & &
\nonumber
\end{eqnarray}
\[
\delta _{u,d}=0,
\]
\begin{eqnarray}  \label{(3)}
\forall u\in V, & &
\quad\quad\quad\quad\quad\quad\quad\quad\quad\quad\quad\quad\quad\quad
\end{eqnarray}
\[
\left\vert E_{u}\right\vert +\sum\limits_{e_{u,v}\in \overline{E}_{u}}\alpha
_{u,v}=\sum\limits_{d\in D}d\delta _{u,d},
\]
\begin{eqnarray}  \label{(4)}
\forall u\in V,\forall d\in D, & & \quad\quad\
\quad\quad\quad\quad\quad\quad\quad\quad
\end{eqnarray}
\[
\delta _{u,d}\leq \theta _{c_{u},d},
\]
\begin{eqnarray}  \label{(5)}
\forall c\in C,\forall d\in D, & & \quad\quad\
\quad\quad\quad\quad\quad\quad\quad\quad
\end{eqnarray}
\[
\theta _{c,d}\leq \sum\limits_{u\in V:c_{u}=c}\delta _{u,d},
\]
\begin{eqnarray}  \label{(6)}
\forall c\in C,\forall d\in D, & & \quad\quad\
\quad\quad\quad\quad\quad\quad\quad\quad
\end{eqnarray}
\[
\left( k-1\right) \theta _{c,d}\leq \sum\limits_{\overline{c}\in C:\overline{%
c}\neq c}\theta _{\overline{c},d}.
\]

Constraint (\ref{(1)}) ensures that the degree of each vertex is unique, and
constraint (\ref{(2)}) prunes unnecessary candidate degrees for each vertex.
The degree for each vertex $u$ must be no smaller than the number of
originally incident edges. In addition, it cannot exceed the sum of the
number of originally incident edges and the number of adding edge
candidates. The left-hand-side of constraint (\ref{(3)}) represents the
degree of vertex $u$, and constraint (\ref{(1)}) guarantees that $\delta
_{u,d}$ is $1$ for only a single $d$. In this way, constraint (\ref{(3)})
together with constraint (\ref{(1)}) ensure that binary variable $\delta
_{u,d}$ can find the correct degree of each vertex.

Constraints (\ref{(4)}) and (\ref{(5)}) collect the degrees of the vertices
in each community. If the degree value of vertex $u$ is $p$, i.e., $\delta
_{u,p}=1$, then constraint (\ref{(4)}) states that the corresponding
community must have at least one vertex with the degree as $p$, i.e., $%
\theta _{c_{u},p}=1$. In contrast, for any other degree value $q$, $q\neq $ $%
p$, constraints (\ref{(1)})-(\ref{(3)}) ensure that $\delta _{u,q}=0$ must
hold. In this case, $0\leq \theta _{c_{u},q}$ must be true when $\theta
_{c_{u},q}$ is either $0$ or $1$. Note that this constraint does not limit
the value of $\theta _{c_{u},d}$ in this case. However, if the degree value
of every vertex $u$ in community $c$ is not $q$, i.e., $\delta _{u,q}=0$,
then the right-hand-side of constraint (\ref{(5)}) is $0 $ and thereby
ensures that $\theta _{c,d}$ in the left-hand-side must be $0$. Therefore,
constraints (\ref{(4)}) and (\ref{(5)}) ensure that binary variable $\theta
_{c,d}$ can find and represent the degrees of the vertices in each community.

Constraint (\ref{(6)}) implements the $k$-structural diversity.
Specifically, if community $c$ has at least one vertex with the degree $d$,
i.e., $\theta _{c,d}=1 $, then this constraint guarantees that there must
exist at least $k-1$ other communities, where each of them also has a vertex
with the degree as $d $. In this case, for each community $\overline{c}$
with $\theta _{\overline{c},d}$ as $1$, constraint (\ref{(5)}) will assign
the degree of at least one vertex $u$ in community $\overline{c}$ to be $d$,
and constraint (\ref{(3)}) will then add several edges to $u$ to fulfill the
degree requirement. Therefore, constraint (\ref{(6)}) is able to achieve the
$k$-structural diversity in $k$-SDA.


\subsection{Formulation with Splitting Vertex as well}

\label{IP-2} 

\begin{table}[tbp] \centering%
\caption{The decision variables of $k$-SDA.}%
\begin{tabular}{l|l}
\hline\hline
Notation & Description \\ \hline
$\alpha _{u,v,i,j}$ & binary variable; $\alpha _{u,v,i,j}=1$ if an edge is
added \\
& to connect substitute vertex $i$ of $u$ and $j$ of $v$; \\
& otherwise, $\alpha _{u,v,i,j}=0$, $u\in V$, $e_{u,v}\in \overline{E}_{u}$,
\\
& $i\in S_{u}$, $j\in S_{v}$ \\
$\beta _{u,i,j}$ & binary variable; $\beta _{u,i,j}=1$ if an edge is added
\\
& to connect the substitute vertices $i$ and $j$ of $u$; \\
& otherwise, $\beta _{u,i,j}=0$, $u\in V$, $i,j\in S_{u}$, $i\neq j$ \\
$\eta _{u,v,i,j}$ & binary variable; $\eta _{u,v,i,j}=1$ if the original edge
\\
& $e_{u,v}$ connects the substitute vertex $i$ of $u$ and $j$ \\
& of $v$; otherwise, $\eta _{u,v,i,j}=0$, $u\in V$, $e_{u,v}\in E_{u}$, \\
& $i\in S_{u}$, $j\in S_{v}$ \\
$\pi _{u,i}$ & binary variable; $\pi _{u,i}=1$ if the substitute \\
& vertex $i$ of $u$ is active; otherwise, $\pi _{u,i}=0$, \\
& $u\in V$, $i\in S_{u}$ \\
$\delta _{u,i,d}$ & binary variable; $\delta _{u,i,d}=1$ if the degree of \\
& substitute vertex $i$ of $u$ is $d$; otherwise, \\
& $\delta _{u,i,d}=0$, $u\in V$, $i\in S_{u}$, $d\in D$ \\
$\theta _{c,d}$ & binary variable; $\theta _{c,d}=1$ if there exists at \\
& least one vertex in $c$ with its degree as $d$, \\
& $c\in C$, $d\in D$ \\ \hline\hline
\end{tabular}%
\label{Split Variable}%
\end{table}%

We now extend the Integer Programming formulation in Section %
\ref{IP-1} to consider both operations in $k$-SDA. Table \ref{Split Variable}
shows the modified decision variables, where subscripts for substitute
vertices are included in variables $\alpha _{u,v,i,j}$ and $\delta _{u,i,d}$%
. To ensure that each substitute vertex in $S_{v}$ has at least one incident
edge in $E_{v}$, we incorporate variable $\eta _{u,v,i,j}$ to assign the
edges in $E_{v}$ to the substitute vertices, and $\beta _{u,i,j}$ represents
the edges between substitute vertices of $v$. Please note that we do not
enforce that every substitute vertex in $S_{v}$ must have an incident edge.
Instead, our formulation allows some vertices in $S_{v}$ to have no incident
edge. In this case, these vertices are not actually split from $v$, and we
regard these vertices \textit{inactive} in $S_{v}$. In the extreme case, if
only one vertex in $S_{v}$ is \textit{active} and has incident edges, the
vertex represents $v$ in our formulation, and $v$ is actually not split in $%
k $-SDA. In our formulation, to avoid missing the globally optimal
solutions, $S_v$ has a sufficient number of candidate substitute
vertices, and only active substitute vertices are included or added
to $G$ in the solutions for users.

The objective function of $k$-SDA with both operations is as follows.%
\[
\min \omega \left( -\left\vert V\right\vert +\sum\limits_{u\in
V}\sum\limits_{i\in S_{u}}\pi _{u,i}\right) +\sum\limits_{e_{u,v}\in
\overline{E}}\sum\limits_{i\in S_{u}}\sum\limits_{j\in S_{v}}\alpha
_{u,v,i,j}+\sum\limits_{e_{u,v}\in E}\left[ -1+\sum\limits_{i\in
S_{u}}\sum\limits_{j\in S_{v}}\eta _{u,v,i,j}\right] .
\]%
The first part represents the cost from operation Splitting Vertex,
and note that no cost is incurred if no such operation is performed,
i.e., there is only one active substitute vertex in $S_{u}$ for each
$u$ in $V$. The second and third terms correspond to the cost from
operation Adding Edge. Moreover, the edges between the substitute
vertices of the same vertex, $\beta _{u,i,j}$, induce no cost. The
problem has the following constraints,

\begin{eqnarray} \label{(7)}
\forall u\in V,\forall i\in S_{u},\quad \quad \quad \quad \quad \quad \quad
\quad \quad \quad
\end{eqnarray}%
\[
\sum\limits_{d\in D}\delta _{u,i,d}=1,
\]%
\begin{eqnarray} \label{(8)}
\forall u\in V,\forall i\in S_{u},\quad \quad \quad \quad \quad
\quad \quad \quad \quad \quad
\end{eqnarray}%
\[
\sum\limits_{e_{u,v}\in E_{u}}\sum\limits_{j\in S_{v}}\eta
_{u,v,i,j}+\sum\limits_{j\in S_{u}:i\neq j}\beta
_{u,i,j}+\sum\limits_{e_{u,v}\in \overline{E}_{u}}\sum\limits_{j\in
S_{v}}\alpha _{u,v,i,j}
\]%
\[
=\sum\limits_{d\in D}d\delta _{u,i,d},
\]%
%
\begin{eqnarray} \label{(9)}
\forall u\in V,\forall i\in S_{u},\forall d\in D,\quad \quad \quad
\quad \quad \quad
\end{eqnarray}%
%
\[
\delta _{u,i,d}\leq \theta _{c_{u},d},
\]%
%
\begin{eqnarray} \label{(10)}
\forall c\in C,\forall d\in D,\quad \quad \quad \quad \quad \quad
\quad \quad \quad \quad
\end{eqnarray}%
%
\[
\theta _{c,d}\leq \sum\limits_{u\in V:c_{u}=c}\sum\limits_{i\in
S_{u}}\delta _{u,i,d},
\]%
%
\begin{eqnarray} \label{(11)}
\forall c\in C,\forall d\in D,\quad \quad \quad \quad \quad \quad
\quad \quad \quad \quad
\end{eqnarray}%
%
\[
\left( k-1\right) \theta _{c,d}\leq \sum\limits_{\overline{c}\in C:\overline{%
c}\neq c}\theta _{\overline{c},d},
\]%
%
\begin{eqnarray} \label{(12)}
\forall e_{u,v}\in E,\quad \quad \ \quad \quad \quad \quad \quad
\quad \quad \quad \quad \quad
\end{eqnarray}%
%
\[
\sum\limits_{i\in S_{u}}\sum\limits_{j\in S_{v}}\eta _{u,v,i,j}\geq
1,
\]%
%
\begin{eqnarray} \label{(13)}
\forall u\in V,\forall e_{u,v}\in E_{u},\forall i\in S_{u},\forall
j\in S_{v},\quad
\end{eqnarray}%
%
\[
\eta _{u,v,i,j}\leq \pi _{u,i},
\]%
%
\begin{eqnarray} \label{(15)}
\forall u\in V,\forall e_{u,v}\in \overline{E}_{u},\forall i\in
S_{u},\forall j\in S_{v},\quad
\end{eqnarray}%
%
\[
\alpha _{u,v,i,j}\leq \pi _{u,i},
\]%
%
\begin{eqnarray} \label{(16)}
\forall u\in V,i\in S_{u},\forall j\in S_{u},i\neq j,\quad \quad
\quad
\end{eqnarray}%
%
\[
\beta _{u,i,j}\leq \pi _{u,i}. \]%
Constraints (\ref{(7)}), (\ref{(8)}), (\ref{(9)}), (\ref{(10)}), and (\ref%
{(11)}) are similar to constraints (\ref{(1)}), (\ref{(3)}), (\ref{(4)}), (%
\ref{(5)}), and (\ref{(6)}). The first term in constraint (\ref{(8)}) is
different from the one in (\ref{(3)}), in which every original edge in $E$
is connected to vertex $u$. In contrast, here we allow the edges in $E_{u}$
to be distributed to the substitute vertices of $u$, while more edges are
also allowed to be added. The left-hand-side of (\ref{(8)}) thereby finds
the degree of each substitute vertex $i$ of $u$.

Constraints (\ref{(12)})-(\ref{(15)}) allocate the original edges in
$E$ to substitute vertices, add more edges, and identify the
corresponding active substitute vertices. Constraint (\ref{(12)})
ensures that each original edge connecting vertices $u$ and $v$ in
$k$-SDA must connect a substitute vertex of $u$ and a substitute
vertex of $v$ here, while new edges are also allowed to be added.
Constraints (\ref{(13)}), (\ref{(15)}), and (\ref{(16)}) guarantee
that a substitute vertex is active when the vertex has at least one
incident edge.

\section{Scalable Approaches}

In this section, we solve the $k$-SDA problem on large scale social
networks. Anonymization of large scale social networks with minimal
information distortion is always challenging because directly enumerating
possible solutions is computationally infeasible. Heuristically,
anonymization problems can be solved by a one-step framework which directly
adjusts a graph to satisfy the privacy requirements \cite%
{k-iso,k-neighbor,k-automor}, or by a two-step framework consisting of
degree sequence anonymization and graph re-construction subjected to
anonymized degree sequence \cite{k-deg}. For $k$-SDA, note that the degree
sequence in the first step presents limited structural information due to
the dimension incurred from the community information, while deriving
additional information in the first step is so computationally intensive
that an algorithm becomes less scalable. Therefore, in this paper, we design
the algorithms to solve the $k$-SDA problem based on the one-step framework.

To ensure good scalability and achieve the anonymization with minimal
information distortion, we propose four algorithms based on the following
concepts. First, our algorithms anonymize the vertices one-by-one such that
the graph anonymization can be efficiently achieved with only one scan of
the vertices. Second, to efficiently minimize the total anonymization cost,
we anonymize the vertices in orders of degrees and handle a set of vertices
with similar degrees to avoid searching for a large amount of combinations.
Third, to consider the community information, we propose two procedures,
\textsc{creation} and \textsc{mergence}, to anonymize each vertex $v$
efficiently. Specifically, \textsc{creation} forms a new anonymous group for
protecting $v$, such that other similar vertices that have not been
considered can be anonymized via this new group and share the same degree
with $v$. In addition to creating new anonymous groups for anonymization,
\textsc{mergence} lets $v$ join an existing anonymous group if joining the
group only incurs a small anonymization cost. Consequently, the above two
procedures enable each vertex to be anonymized efficiently, and the graph
anonymization can thereby be achieved with minimal information distortion.

In this paper, we propose four algorithms to solve $k$-SDA. The first
algorithm, named \textbf{EdgeConnect}, specially aims at minimizing
information distortion. That is, EdgeConnect applies operation Adding Edge
alone since adding edges within a community does not destroy existing
semantic information, such as friendships, and makes limited changes over
the whole graph. It should be noted that, with sole use of Adding Edge, the
degrees of vertices can only increase. EdgeConnect thus considers the
vertices in \textit{decreasing order} of the degrees to first anonymize the
vertices with large degrees, so that we have more chances to achieve the
anonymization of subsequent vertices without affecting existing anonymous
groups. Second, to provide more variety for anonymization, we then extend
EdgeConnect with operation Splitting Vertex and propose the \textbf{%
CreateBySplit} algorithm. CreateBySplit utilizes the same anonymization flow
as EdgeConnect, but leverages Splitting Vertex if the anonymization cannot
be achieved by Adding Edge alone. Incorporating Splitting Vertex can not
only provide more chances to achieve the anonymization but also incur less
information distortion. Differing from the previous two algorithms, which
focus on minimizing the information distortion, the third algorithm, named
\textbf{MergeBySplit}, is designed to guarantee the anonymization for the
social networks that are difficult to be anonymized with respect to a high
privacy level $k$. For this purpose, MergeBySplit anonymizes the vertices in
\textit{increasing order} of the degrees, and the creation of new anonymous
groups with small degrees thereby allows us to protect a vertex with any
larger degree by operation Splitting Vertex. Finally, we propose the fourth
algorithm, named \textbf{FlexSplit}, to improve Algorithm MergeBySplit and
reduce the number of generated substitute vertices in the objective function
of $k$-SDA. Specifically, in addition to anonymizing a vertex by splitting
it into members of the existing anonymous groups
as in Algorithm MergeBySplit, FlexSplit is endowed with a new splitting
strategy, which splits a group of vertices to generate a new anonymous group
of a target degree for anonymization. With the capability of looking forward
$k$ subsequence vertices for anonymization, FlexSplit is able to reduce the
substitute vertices with the two splitting strategies. FlexSplit is thus
more flexible and preserves more data utilities than MergeBySplit under the
same guarantee of anonymization. 

Before we introduce these algorithms in detail, we first define the
anonymous group, which considers not only the number of vertices of the same
degree but also the distribution of the vertices over the
communities.\smallskip

\textbf{Definition 2.} An anonymous group of degree $d$, denoted as $g_{d}$,
consists of the vertices with degree $d$, i.e., $g_{d}=\{v|d_{v}=d\}$. A $%
g_{d}$ is a $k$-SDA group, denoted as $\widehat{g}_{d}$, if $%
C_{g_{d}}=\{c_{v}|v\in g_{d}\}$ and the cardinality of $C_{g_{d}}$ is no
smaller than $k$, i.e., $|C_{g_{d}}|\geq k$.\smallskip

\textbf{Lemma 1.} If every vertex $v$ in $G(V,E,C)$ belongs to a $k$-SDA
group, $G(V,E,C)$ must satisfy $k$-SDA.\smallskip

Given a graph $G(V,E,C)$, the objective is to assign every vertex $v$ to a
group $\widehat{g}_{d}$ with minimal information distortion. In the next
sections, we present the details of our algorithms.

\vspace{-2mm}

\subsection{Algorithm EdgeConnect}

The EdgeConnect algorithm is designed for minimizing information distortion
on large-scale graphs. For this purpose, the EdgeConnect algorithm
incorporates operation Adding Edge to anonymize the vertices one-by-one in
decreasing order of their degrees to avoid enumerating all possible
combinations, which is computationally infeasible. One merit of EdgeConnect
is that the existing information is never removed, and the added local new
edges within each community incur few changes to the whole graph. Moreover,
procedures \textsc{creation} and \textsc{mergence} are utilized in this
algorithm, and any existing $k$-SDA group is never removed in order to avoid
re-anonymizing the vertices and increasing the computation cost. As a
result, EdgeConnect has very good scalability, which is shown in our
experiments.

\begin{figure}[tbh]
\begin{center}
$%
\begin{array}{ccc}
\includegraphics[width=0.22\textwidth]{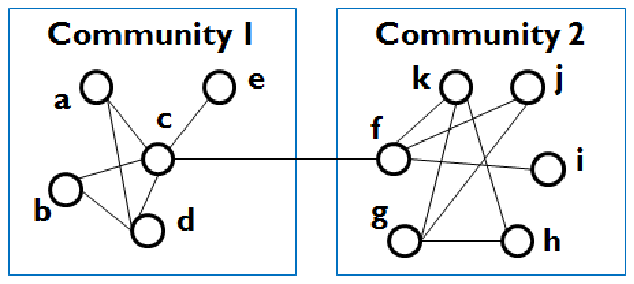} &  & %
\includegraphics[width=0.22\textwidth]{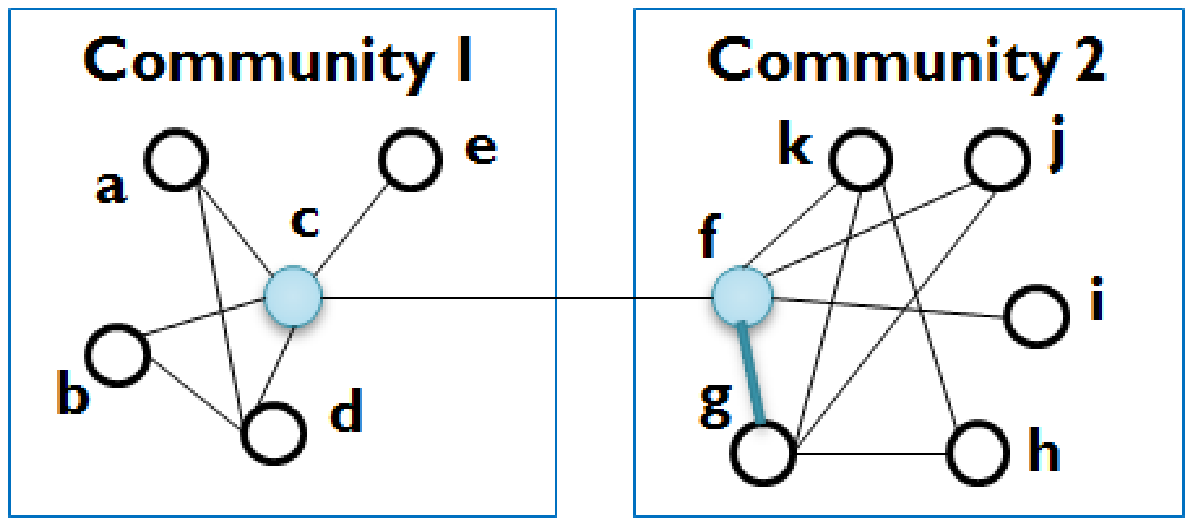} \\
{(a)} &  & {(b)} \\
\includegraphics[width=0.22\textwidth]{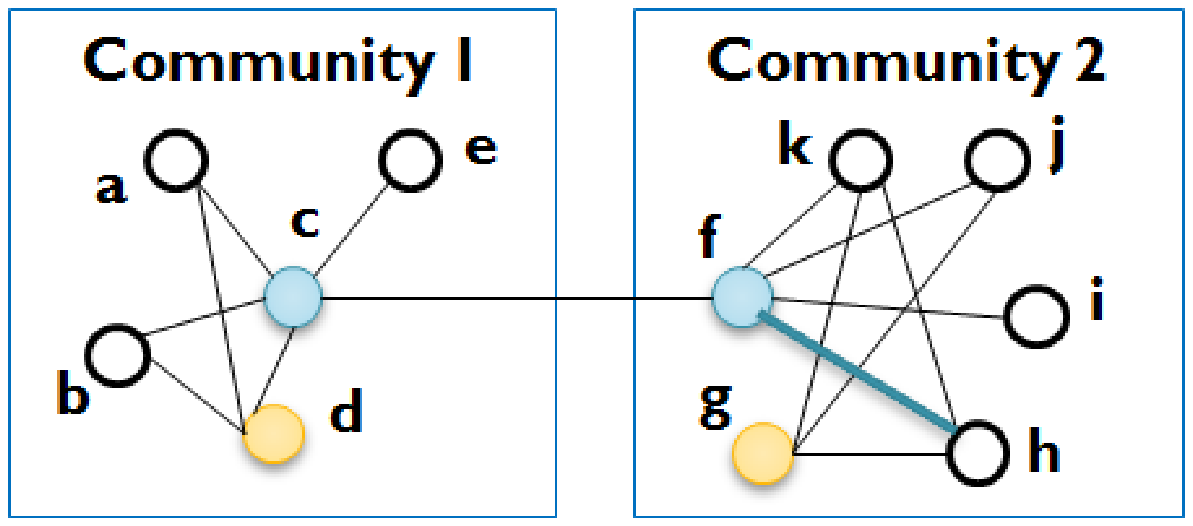} &  & %
\includegraphics[width=0.22\textwidth]{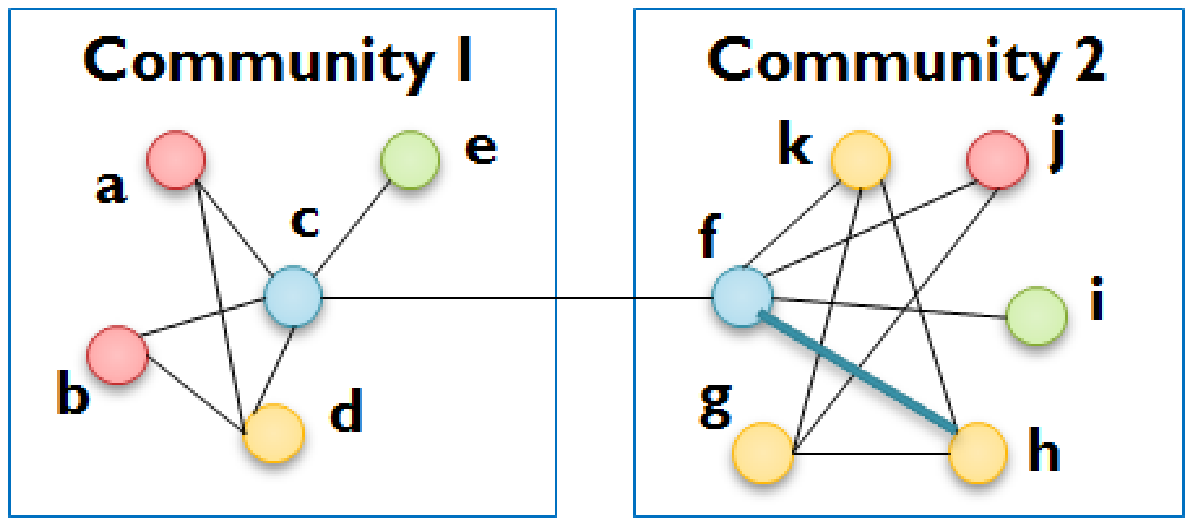} \\
{(c)} &  & {(d)}%
\end{array}
$%
\end{center}
\par
\vspace{-2mm} \caption{Example of anonymization by EdgeConnect.}
\label{EC}
\end{figure}

The rationale of Algorithm EdgeConnect is to adjust the vertex degrees
one-by-one with operation Adding Edge in order to let every vertex share the
same degree with other vertices in at least $k$ different communities. To
avoid examining all possibilities, the anonymization begins from a
not-yet-anonymized vertex $v$ of the \textit{largest} degree, since the
power-law degree distribution demonstrated in the previous social network
analysis indicates that each large degree has fewer vertices required to be
anonymized. For a chosen $v$, EdgeConnect utilizes procedure \textsc{mergence%
} and \textsc{creation} to explore the way to anonymize $v$ with minimal
number of new edges. Procedure \textsc{mergence} aims at adjusting the
degree for a vertex $v$ to join an existing $k$-SDA group, while \textsc{%
creation} is designed to collaborate with other not-yet-anonymized vertices
to generate a new $k$-SDA group with a new degree. In the example of Figure %
\ref{EC}(a), the first vertex to be anonymized is vertex $c$ because its
degree is the largest one. At the beginning, procedure \textsc{mergence} is
unable to anonymize $c$ since no $k$-SDA group has been generated, and
procedure \textsc{creation} thus generates a new anonymous group of degree 5
by adding an edge connecting $f$ and another vertex in the same community,
such as $g$. At this point, the new $k$-SDA group is \{$c$, $f$\} as shown
in Figure \ref{EC}(b). EdgeConnect repeats the above process until all the
vertices are successfully anonymized.

The details of each step are presented as follows. First, procedure \textsc{%
mergence} protects a vertex $v$ with an existing $k$-SDA group
$g_{d}$. As all vertices in $k$-SDA group $g_{d}$ share the same
degree $d$ for structural diversity, the cost for $v$ to be
anonymized (by the operation Adding Edge) in $g_{d}$ is
\begin{equation}
Cost_{\mbox{\textsc{mrg}}}(v,d)=\left\{
\begin{array}{ll}
d-d_{v}, & \mbox{if }d\geq d_{v} \\
\infty , & \mbox{otherwise.}%
\end{array}%
\right.  \label{eq:cm0}
\end{equation}%
The minimal \textsc{mergence} cost for $v$ is evaluated as $\min {}_{%
\widehat{d}}$ $Cost_{\mbox{\textsc{mrg}}}(v,\widehat{d})$ to find a suitable
$k$-SDA group for $v$ from all existing $k$-SDA groups, where $\widehat{d}$
is the degree of a $k$-SDA group $\widehat{g}_{d}$. For example, if there
are three existing $k$-SDA groups with degrees 2, 5 and 6, the minimal
\textsc{mergence} cost for a vertex $v$ of degree 4 is 1 by increasing its
degree to $\widehat{d}=5$. Next, for procedure \textsc{creation}, which
introduces a new $k$-SDA group, our algorithm finds the vertices distributed
in other $k-1$ communities to join this new group. Specifically, the
diversity of a group $g_{d}$ is first defined as

\noindent
\begin{equation}
Div(g_{d_{v}})=\left\{
\begin{array}{ll}
1, & \mbox{if }|C_{g_{d_{v}}}|\geq k \\
\infty , & \mbox{if }|C_{g_{d_{v}}}|<k,%
\end{array}%
\right.  \label{eq:ds}
\end{equation}%
where $C_{g_{d_{v}}}=\{c_{u}|u\in g_{d_{v}}\}$. Accordingly, the minimal
cost for $v$ in \textsc{creation}\ is
\begin{equation}
\begin{array}{l}
Cost_{\mbox{\textsc{crt}}}(v)= \min_{U}\{Div(U)\times \sum_{u\in U}Cost_{%
\mbox{\textsc{mrg}}}(u,d_{v})\},\quad%
\end{array}
\label{eq:cc0}
\end{equation}%
where $U$ is any subset of $k$ vertices that have not been anonymized,
including $v$. For example, if $k$ is 2 and not-yet-anonymized vertices $v$
and $u$ in different communities are of degrees 4 and 2, respectively, when $%
g_{4}$ has not been previously generated, a simple way for anonymizing $v$
is to create a new $k$-SDA group $g_{4}=\{v,u\}$ by increasing the degree of
$u$ to 4. However, to avoid exploring every possible $U$, we sort all
not-yet-anonymized vertices of each community in the decreasing order of
their degrees, and the vertex with the largest degree in each community is
chosen for $U$ since the degree difference between those vertices and $v$ is
the smallest. If $|C|>k$, only $k$ of the above vertices with the largest
degrees are selected to construct $U$ such that $|U|=k$. 
Therefore, finding the anonymization costs for each vertex $v$ is
computationally efficient.

In our algorithm design, the not-yet-anonymized vertices in each community
are sorted in the decreasing order of their degrees. Let $s_{c}$ denote the
order set of the vertices for community $c$, and $s_{c}(i)$ be the vertex
with the $i$-th largest degree in $c$. We anonymize the vertices one-by-one
with \textsc{mergence} and \textsc{creation} as follows. We first choose the
largest degree vertex $v$ among $s_{1}(1),\ldots ,s_{|C|}(1)$. If $\min {}_{%
\widehat{d}}$ $Cost_{\mbox{\textsc{mrg}}}(v,\widehat{d})$ is smaller than $%
Cost_{\mbox{\textsc{crt}}}(v)$, procedure \textsc{mergence} increases the
degree of $v$ by adding ($\widehat{d}-d_{v}$) edges connecting $v$ and the ($%
\widehat{d}-d_{v}$) subsequent vertices, which are not yet connected to $v$,
in $s_{c_{v}}$. We then update $s_{c_{v}}$, and note that the update of $%
s_{c_{v}}$ is efficient given that only ($\widehat{d}-d_{v}$) vertices
increase their degrees by 1. Otherwise, procedure \textsc{creation} finds $U$%
, increases the degree of each vertex $u$ in $U$ to $d_{v}$ in the same way,
and updates the corresponding $s_{c_{u}}$ as well. We present the proceeding
illustrative example.

\textbf{Example 1.} Consider the graph in Figure \ref{EC}(a) with
$k$ as 2. In the decreasing order of the degrees, the vertex orders
are $s_{1}=cdabe $ and $s_{2}=fgkhji$. Accordingly, the first
considered vertex (the largest degree vertex) is $c$. From Formula
(\ref{eq:cm0}), the \textsc{mergence}
cost for $c$ is infinity as there is no 2-SDA group. According to Formula (%
\ref{eq:cc0}), the \textsc{creation} cost for $c$ is 1, and the set
$U$ corresponding to the minimal cost consists of $c$ and $f$ (the
first vertex in $s_{c}$). Therefore, vertex $c$ is anonymized by
\textsc{creation} and an edge is added between $f$ and $g$.
Consequently, a new 2-SDA group of degree
5 is generated, and the vertex orders are updated to $s_{1}=dabe$ and $%
s_{2}=gkhji$. Figure \ref{EC}(b) shows the result after this iteration,
where the anonymized vertices are shaded. $\blacksquare $

The above two procedures can anonymize every vertex with a minimal cost at
each iteration. However, 
Since adding an edge increases the degrees of two vertices, the
newly added edge ($f$,$g$) in Figure \ref{EC} not only increases the
degree of vertex $f$ for creating a 2-SDA group of degree 5 but also
increases the degree of vertex $g$ simultaneously. Nevertheless,
this increment of the degree on $g$ incurs additional cost to
anonymize the not-yet-anonymized vertex $g$.
To avoid the above case, we define redirectable edges and propose
edge-redirection operation, so that edge ($f$,$g$) can be properly
replaced by another edge, such as ($f$,$h$), without revoking the
anonymization of vertices $c$ and $f$ examined previously.

\smallskip \textbf{Definition 3.} An added edge $(w,v)$, where $w$ is an
anonymized vertex and $v$ is a not-yet-anonymized vertex in the same
community, is said redirectable away from $v$ if there is another
not-yet-anonymized vertex $x$ in the same community not yet
connecting to $w$.
Defined on such an edge, the edge-redirection operation performs
\[
\widehat{E}\leftarrow \widehat{E}/(w,v)\cup (w,x)\mbox{,}
\]%
where $\widehat{E}$ is the set of existing added edges. 
\smallskip


Let $R_{v}$ denote the set of edges that are redirectable away from
$v$. The edge-redirection operation allows us to reduce the degree
of $v$ without changing the degree of any vertex $w$ that has been
anonymized in a $k$-SDA group.
Therefore, we can modify procedure \textsc{mergence} in the following way to
allow $v$ to join the group with a smaller degree, by redirecting some added
edges incident to $v$.
\begin{equation}
Cost_{\mbox{\textsc{mrg}}}(v,d)=\left\{
\begin{array}{ll}
0, & \mbox{if }d_{v}\geq d\geq d_{v}-|R_{v}| \\
d-d_{v}, & \mbox{if }d>d_{v} \\
\infty , & \mbox{otherwise.}%
\end{array}%
\right.  \label{eq:cm1}
\end{equation}%
Thus, to find a suitable $k$-SDA group, we derive the minimal \textsc{%
mergence} cost for $v$ as $\min {}_{\widehat{d}}$ $Cost_{\mbox{\textsc{mrg}}%
}(v,\widehat{d})$, where $\widehat{d}$ is the degree of a $k$-SDA group $%
\widehat{g}_{d}$. Similarly, we modify procedure \textsc{creation} and
derive the minimal cost of creating a new $k$-SDA group for $v$ as
\begin{equation}  \label{eq:cc1}
\begin{array}{l}
Cost_{\mbox{\textsc{crt}}}(v) = \min_{U} \{Div(U) \times \sum_{u\in U}Cost_{%
\mbox{\textsc{mrg}}}(u,d_{v}-|R_{v}|)\},%
\end{array}%
\end{equation}
\noindent where $U$ is any subset of $k$ vertices that have not been
anonymized, including $v$. As a result, with the edge-redirection
operation and the two modified procedures, we are able to reuse the
edges added previously to further reduce the anonymization cost.

\begin{figure}[tbh]
\begin{center}
$%
\begin{array}{c}
\includegraphics[width=0.45\textwidth]{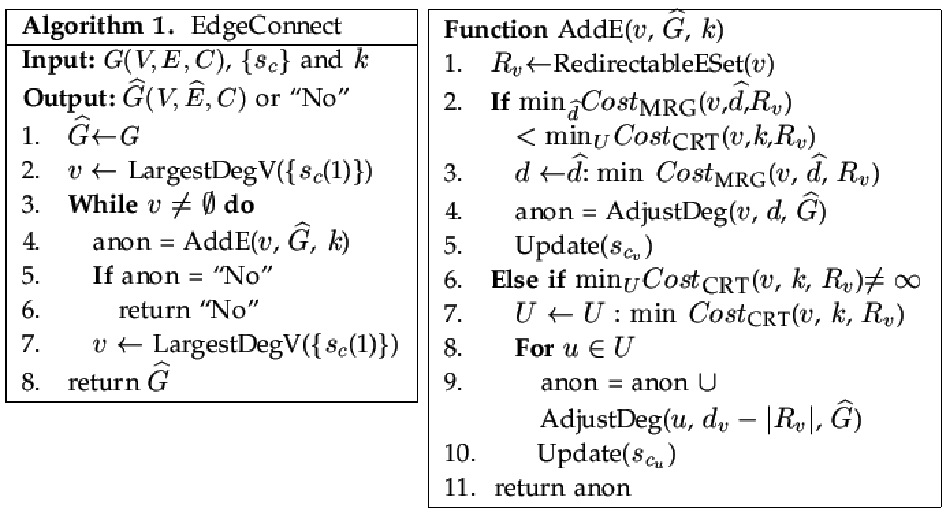}%
\end{array}
$%
\end{center}
\par
\vspace{-2mm} \caption{The pseudo code of EdgeConnect.}
\label{code-ec}
\end{figure}

In the following, we propose Algorithm EdgeConnect (Algorithm 1 in Figure %
\ref{code-ec}) based on the modified \textsc{mergence} and \textsc{creation}%
. For each vertex $v$, EdgeConnect first finds the set $R_{v}$ of added
edges that can be redirected away from $v$. More specifically,
$R_{v}$ is a subset of new edges incident to $v$ added during operation
Adding Edge. For every edge $(w,v)$ in $R_{v}$, there must exist a vertex $x$
in the same community of $v$ such that $x$ shares no edge with the
anonymized $w$. To calculate $R_{v}$ efficiently, Algorithm EdgeConnect
examines every new edge $(w,v)$ incident to $v$ to find $V_{C_{v}}-N_{w}$,
where $V_{C_{v}}$ is the set of not-yet-anonymized vertices in the same
community of $v$, and $N_{w}$ is the set of neighboring vertices of $w$. We
add $(w,v)$ to $R_{v}$ if $V_{C_{v}}-N_{w}$ is not an empty set. For vertex $%
g$ in Figure \ref{EC}(b) following Example 1, $(f,g)$ is in $R_{g}$ since $%
\{g,h,i,j,k\}-\{g,i,j,k\}\neq \emptyset $. In Community 2, there is a vertex
$h$ that does not connect to the anonymized vertex $f$. 
After identifying $R_{v}$, the costs induced from \textsc{mergence} and
\textsc{creation} for $v$ are evaluated by (\ref{eq:cm1}) and (\ref{eq:cc1}%
). If the \textsc{mergence} cost is smaller than the \textsc{creation} cost,
the degree of $v$\ is increased by Adding Edge or decreased by the
edge-redirection operation. Otherwise, EdgeConnect anonymizes $v$ by
creating a new $k$-SDA group with the vertices in $U$ that minimizes the
cost in (\ref{eq:cc1}). EdgeConnect returns the anonymized graph $\widehat{G}%
(V,E\cup \widehat{E},C)$ and obtains the anonymization cost. \smallskip

\textbf{Example 2.} We continue the example in Figure \ref{EC}. However,
procedures \textsc{mergence} and \textsc{creation} utilize (\ref{eq:cm1})
and (\ref{eq:cc1}) here, instead of (\ref{eq:cm0}) and (\ref{eq:cc0}) as in
Example 1. In this case, $c$ is still the first vertex to be anonymized.
However, at the next iteration as shown in Figure \ref{EC}(b), without the
edge-redirection operation, $g$ can only be anonymized by adding another
edge to increase its degree to 5 (by \textsc{mergence}), or by adding an
edge between $d$ and $e$ to create a new 2-SDA group of degree 4 (by \textsc{%
Creation}). In both ways, we need to add an edge to the graph. In contrast,
the edge-redirection operation is able to avoid this additional edge.
Specifically, for vertex $g$, EdgeConnect first finds $R_{g}=\{{(f,g)\}}$.
The \textsc{Creation} cost for $g$ is thus 0, and the set $U$ that minimizes
this cost is $\{{d,g\}}$. The \textsc{mergence} cost for $g$ is 1 because
the only 2-SDA group is of degree 5. Therefore, EdgeConnect anonymizes $g$
by creating a new 2-SDA group consisting of $d$ and $g$, and redirecting the
edge $(f,g)$ to $(f,h)$. Consequently, the edge-redirection operation
enables us to anonymize $g$ with zero cost. Figure \ref{EC}(c) shows the
result after the second iteration of anonymization, where the anonymized
vertices belonging to the same 2-SDA groups are shaded in the same color.
When EdgeConnect terminates, the final anonymous result is shown in Figure %
\ref{EC}(d). $\blacksquare $ \smallskip


\vspace{-2mm}

\subsection{Algorithm CreateBySplit}

In this subsection, we extend Algorithm EdgeConnect with operation Splitting
Vertex and propose Algorithm CreateBySplit. Compared to EdgeConnect,
CreateBySplit is a more realizable solution because Splitting Vertex will
increase the number of vertices in a community and provide a greater number
of chances to achieve the anonymization.

\begin{figure}[tbh]
\begin{center}
$%
\begin{array}{c}
\includegraphics[width=0.38\textwidth]{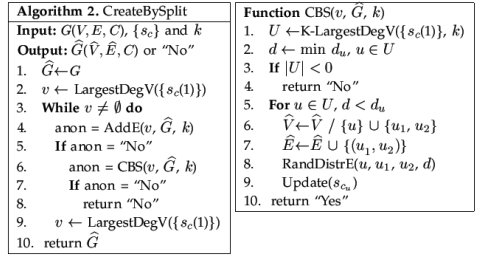}%
\end{array}
$%
\end{center}
\par
\vspace{-2mm} \caption{The pseudo code of CreateBySplit.}
\label{code-cbs}
\end{figure}
\begin{figure}[h]
\centering
\includegraphics[width=0.40\textwidth]{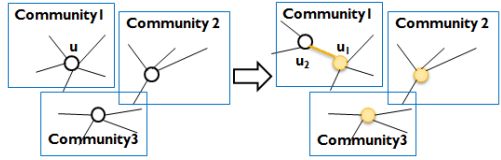} \vspace{-2mm}
\caption{Example of splitting strategy of CreateBySplit.}
\label{cbssplit}
\end{figure}

Specifically, Splitting Vertex replaces a vertex $v$ with a set $S_{v}$ of
substitute vertices, and redistributes incident edges of $v$ to substitute
vertices so that each substitute vertex presents partial truths of $v$.
Splitting Vertex will thus increase the number of vertices and incur higher
information distortion than Adding Edge. To minimize the information
distortion, Splitting Vertex is always regarded as the second choice and
will be applied only if Adding Edge is not able to anonymize the social
network.

In addition, to avoid creating too many vertices and increasing information
distortion, we always use two substitute vertices $v_{1}$ and $v_{2}$ to
replace $v$, and connect $v_{1}$ and $v_{2}$ with an edge. This approach can
limit the incrementation of the length for the shortest path between any
pair of vertices due to the split of a vertex.


In other words, when Adding Edge is not able to anonymize the social
network (Algorithm 2 in Figure \ref{code-cbs}), CreateBySplit
anonymizes a given vertex $v$ with Splitting Vertex in the following
way.
Let $U$ denote the vertex set consisting of $k$ not-yet-anonymized
vertices with the largest degrees in $k$ different communities.
CreateBySplit generates a new $k$-SDA of degree $d$ in the following
steps, where $d$ is the maximal degree satisfying $d\leq d_{u}$ for
every $u\in U$. When $d_{u}>d>2$, CreateBySplit (1) replaces $u$
with two substitute vertices $u_{1}$ of degree $d_{u_{1}}=d-1$ and
$u_{2}$ of degree $d_{u_{2}}=d_{u}-d+1$, and then (2) connects
$u_{1}$ and $u_{2}$ with an additional edge $(u_{1},u_{2})$, so that
$d_{u_{1}}=d$ and $d_{u_{2}}=d_{u}-d+2$ eventually. In the 2nd step,
the edge $(u_{1},u_{2})$ is added not only to ensure $d_{u_{1}}=d$
but also reduce the information distortion such as the split of
connected components and the impact in the shortest paths (and their
lengths). On the other hand, when $d_{u}>d=2$, connecting $u_{1}$
and $u_{2}$ with an additional edge $(u_{1},u_{2})$ in the 2nd step
will enforce $d_{u_{2}}=d_{u}-2+2=d_{u_{2}}$ and thus make $u_2$
just another $u$ of the same degree to be anonymized. Similar
situation occurs for
$d=1$. %
To tackle those special cases with $d\leq 2$, CreateBySplit assigns
$d_{u_{1}}=d$ and $d_{u_{2}}=d_{u}-d$ and no longer connects $u_{1}$
and $u_{2}$ with an additional edge. Consequently, in both general
and special cases, $u_{1}$ will be anonymized in the newly generated
$k$-SDA group of degree $d$, while $u_2$ is a not-yet-anonymized
vertex to be subsequently anonymized as with other vertices.

\subsection{Algorithm MergeBySplit}

Here, we propose Algorithm MergeBySplit for the social networks that are
difficult to be anonymized with respect to a high privacy level $k$. In
CreateBySplit, even though Splitting Vertex can generate vertices to
increase the possibility of anonymization for the social networks, the
algorithm still cannot guarantee finding the solution of every instance of $%
k $-SDA. In contrast, MergeBySplit can anonymize every social network, even
for the most difficult one.

\begin{figure}[tbh]
\begin{center}
$%
\begin{array}{c}
\includegraphics[width=0.38\textwidth]{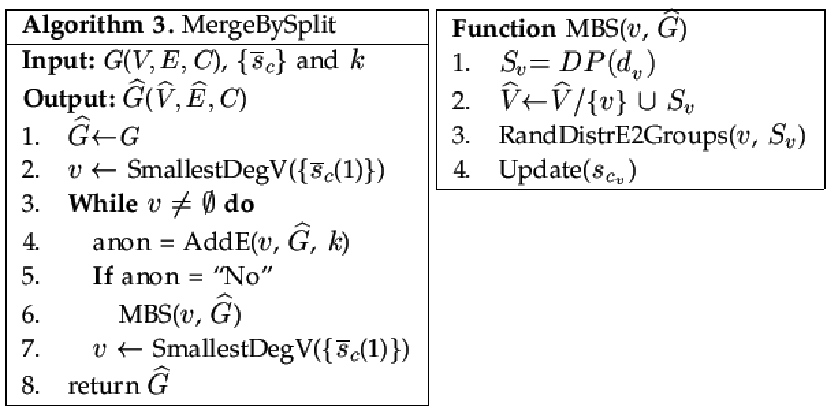}%
\end{array}
$%
\end{center}
\par
\vspace{-2mm} \caption{The pseudo code of MergeBySplit.}
\label{code-mbs}
\end{figure}
\begin{figure}[h]
\label{mbssplit}\centering
\includegraphics[width=0.40\textwidth]{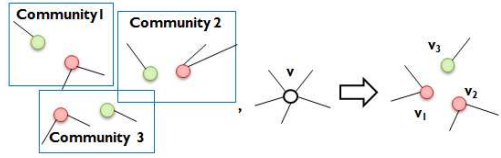}
\caption{Example of splitting strategy of MergeBySplit.}
\vspace{-2mm}
\end{figure}

In more detail, MergeBySplit anonymizes the vertices one-by-one in the
increasing order of the degrees, and performs Splitting Vertex by allowing
each vertex $v$ to be split into more than two substitute vertices protected
by the existing $k$-SDA groups. The rationale of this algorithm is that, the
creation of $k$-SDA groups with small degrees allows us to protect any
vertex $v$ by splitting $v$ into many cohorts of the generated $k$-SDA
groups. In the worst case, we can split a vertex $v$ of degree $d_{v}$ into $%
d_{v}$ substitute vertices of degree 1 to achieve the anonymization for an
arbitrary $k$, $1\leq k\leq |C|$.

However, to reduce the information distortion, when we split a vertex $v$ to
cohorts of the existing $k$-SDA groups, we create the least number of
substitute vertices based on the following dynamic programming.
\begin{equation}  \label{eq:ms}
\vspace{2mm}
\begin{array}{rl}
|S_{v}|= & DP(d_{v}) \\
= & \min \{D(d_{v}), \min {}_{1\leq d < d_{v}}DP(d_{v}-d)+D(d)\},%
\end{array}%
\vspace{1mm}
\end{equation}
where $D(d)=1$, if there is a $k$-SDA group $\widehat{g}_{d}$\ of degree $d$%
; otherwise, $D(d)=\infty $.

We now describe the details of Algorithm MergeBySplit (Algorithm 3 in Figure %
\ref{code-mbs}). MergeBySplit sorts the not-yet-anonymized vertices in each
community in the \textit{increasing order} of the degrees. Let $\overline{s}%
_{c}$ denote the order set of vertices in community c, and $\overline{s}%
_{c}(i)$ be the vertex with the i-th smallest degree. At each iteration, we
anonymize a vertex $v$ with the \textit{smallest degree} $d_{v}$ with
procedures \textsc{mergence} or \textsc{creation} as specified in Algorithm
CreateBySplit. 
If it is too restrictive to anonymize $v$ by Adding Edge and
edge-redirection operations, we perform Splitting Vertex operation to
anonymize $v$. That is, we replace $v$ with a set $S_{v}$ of substitute
vertices as shown in Figure \ref{mbssplit}, i.e.,%
\[
V\leftarrow V/\{{v\}\cup S}_{{v}}{,}
\]

\noindent where the size of $S_{v}$\ is determined by Formula (\ref{eq:ms}%
). Afterward, the edges incident to $v$ are randomly redistributed
to the substitute vertices $v_{1},v_{2},\ldots ,v_{|s_{v}|}$ such
that each substitute vertex $v_{j}$, $j=1,\ldots ,|s_{v}|$, is a
cohort of some existing $k$-SDA group $\widehat{g}_{d}$, i.e.,
$d_{v_{j}}=d$. As shown above, anonymizing $v$ by Splitting Vertex
in this way can always succeed. When all the vertices belong to
$k$-SDA groups, Algorithm MergeBySplit returns the anonymized graph
$\widehat{G}$. \smallskip

\subsection{Algorithm FlexSplit}

In this subsection, we propose Algorithm FlexSplit that improves
MergeBySplit and preserves more utilities of the social networks under the
same guarantee of anonymization. FlexSplit outperforms MergeBySplit by
introducing a new splitting strategy and the capability of looking forward.

To elaborate, in addition to splitting a vertex into substitutes protected
by the existing anonymous groups as MergeBySplit, FlexSplit is endowed with
a new splitting strategy, which identifies a group of vertices and splits
these vertices to generate a new anonymous group of a target degree. In this
way, the degrees of substitute vertices are not constrained to be the same
as those of the existing anonymous groups. FlexSplit is better able to
preserve the degree distribution by setting a large target degree for the
newly generated anonymous group. Moreover, when splitting a group of
vertices together, FlexSplit introduces new edges to connect the substitute
vertices to effectively prevent the partitioning of connected components in
a social network.





With Vertex Splitting operation, FlexSplit is thus more flexible and is able
to anonymize a selected vertex $v$ in the following strategies, for reducing
the number of generated substitute vertices. The first strategy is \textit{%
Single Splitting}, which splits $v$ into multiple substitute vertices as in
MergeBySplit. Let $S^M_v$ denote the minimal set of substitute vertices
generated by Single Splitting, and $S^M_v$ can be derived by Formula (\ref%
{eq:ms}). 
The second strategy is \textit{Group Splitting}, which identifies a
group of vertices and splits those vertices to generate a new
anonymous group of the target degree for anonymization. To create
minimal number of substitute vertices, this strategy splits each
vertex into at most two substitute vertices. The minimal set
$S_{v}^{C}$ of substitute vertices generated by Group Splitting is
thus determined as
\begin{equation}
S_{v}^{C}=2\times \{u|d_{u}>d_{v},u\in W\}{,}  \label{eq:fs}
\end{equation}%
where $W$ is the vertex set consisting of $k$ not-yet-anonymized
vertices with the smallest degrees in $k$ different communities.
Since each node is split into two substitute nodes, we have a
multiplier of 2 in Formula (\ref{eq:fs}).
One of the substitute vertex is anonymized with the target degree
$d_v$ of the newly generated anonymous group and the other has the
remaining degree $d_u-d_v+2$ with an additional edge added to
connect the two substitute vertices.

Furthermore, FlexSplit is also endowed with the capability of looking
forward, to reduce the number of generated substitute vertices in the
objective function of $k$-SDA. In other words, it should be noted that
Single Splitting usually generates fewer substitute vertices than Group
Splitting, especially when $k$ is large. If we simply compare $|S^C_v|$ and $%
|S^M_v|$ and choose the strategy that introduces fewer substitute vertices
to anonymize each selected vertex $v$, Single Splitting will be performed
most of the time for anonymizing $v$ at each iteration, which may result in
generating more substitute vertices in total after many iterations.

\begin{figure}[tbh]
\begin{center}
$%
\begin{array}{c}
\includegraphics[width=0.45\textwidth]{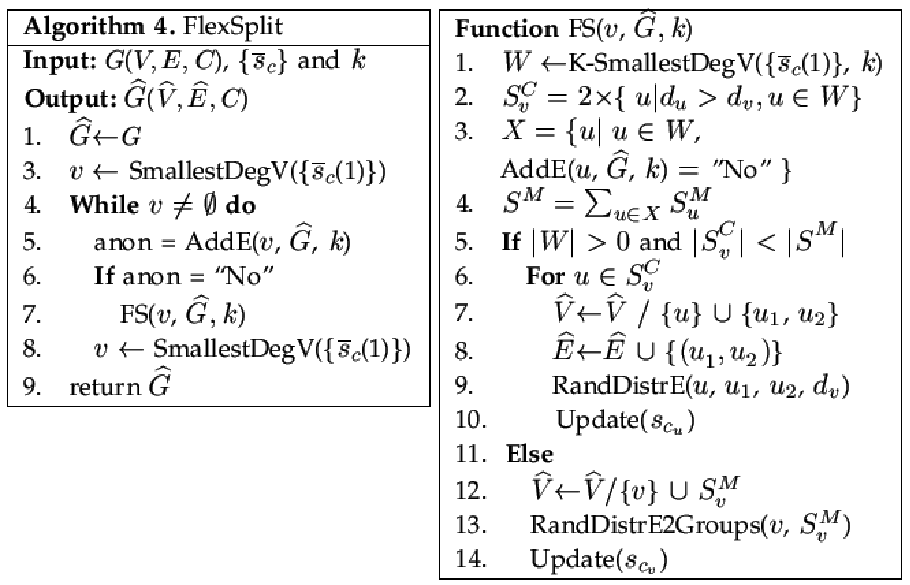}%
\end{array}
$%
\end{center}
\par
\vspace{-2mm} \caption{The pseudo code of FlexSplit.}
\label{code-fs}
\end{figure}

To sidestep this trap, FlexSplit looks forward by identifying, from
$W$, the subset $X$ consisting of the vertices that cannot be
anonymized by Adding Edge alone, and compares the numbers of
substitute vertices $|S^C_v|$ and $\sum_{u \in X} |S^M_u|$, instead
of $|S^C_v|$ and $|S^M_v|$, to choose the splitting strategy.
Specifically, recall that the vertex set involved in procedure
\textsc{Creation} is $W$, and $X$ is the subset of vertices in $W$
such that $X$ cannot be anonymized by Adding Edge alone in both
\textsc{Creation} and the subsequent \textsc{Mergence}. FlexSplit
first examines every vertex of $W$ and initializes $X$ as the set of
vertices that cannot be anonymized by Adding Edge alone in
\textsc{Creation}. Let $u^{\prime }$ denote the vertex of the
largest degree among the vertices that cannot be anonymized in
\textsc{Creation}. $X$ includes the vertices in $W$ whose degrees
are smaller than or equal to $d_{u^{\prime }}$, since the
\textsc{Creation} process of these vertices will also involve
$u^{\prime }$. Afterward, FlexSplit removes some vertices from $X$
such that every remaining vertex in $X$ cannot be anonymized by
Adding Edge alone in \textsc{Mergence}, neither. Let $d_{\max }$
denote the largest degree of the existing $k$-SDA groups. According
to Formula (\ref{eq:cm1}), FlexSplit calculates the
\textsc{Mergence} cost of every vertex $u$ in $X$ with respect to
$d_{\max }$ and removes $u$ from $X$ if $Cost_{MRG}(u,d_{\max
})>|C_{u}|-|N_{u}|-1$, where $C_{u}$ denotes the set of all vertices
in the same community of $u$, and $N_{u}$ represents all the
neighbors of $u$.
After that, FlexSplit compares the numbers of substitute vertices
$|S^C_v|$ and $\sum_{u \in X} |S^M_u|$, and anonymizes $v$ by Group
Splitting if $|S^C_v| < \sum_{u \in X} |S^M_u|$ and by Single
Splitting otherwise.


We now give the complete picture of Algorithm FlexSplit (Algorithm 4 in
Figure \ref{code-fs}). FlexSplit first sorts the not-yet-anonymized vertices
in each community in increasing order of the degrees. Thereafter, at each
iteration, the algorithm tries to anonymize a vertex $v$ of the smallest
degree $d_v$ with procedures \textsc{mergence} and \textsc{creation} as in
MergeBySplit. If it is too restrictive to anonymize $v$ by operations Adding
Edge and edge-redirection, FlexSplit discovers the set $W$ of $k$
not-yet-anonymized vertices with the smallest degrees among all communities,
and computes the minimal set of substitute vertices $S^C_v$ required for
Group Splitting. In addition, FlexSplit also discovers the subset $X$ of $W$
and computes $\sum_{u \in X} |S^M_u|$, where each vertex $u$ in $X$ cannot
be anonymized by Adding Edge. If $|S^C| < \sum_{u \in X} |S^M_u|$, FlexSplit
anonymizes $v$ by Group Splitting. Otherwise, Single Splitting is performed.
When all vertices belong to $k$-SDA groups, FlexSplit returns the anonymized
graph $\widehat{G}$. 


\subsection{Complexity Analysis}

We will now show that the complexities of the four heuristic algorithms. Let
$n$, $m$ and $l$ denote the numbers of vertices, edges and communities of
the input graph $G$, and $d_{\max }$ represents the largest vertex degree in
$G$, $d_{\max }\leq n$.


We derive the space complexity of the four heuristics as follows. First,
storing the whole input graph requires $O(n+m)$ space. For each of the four
heuristics, maintaining a sorted list of not-yet-anonymized vertices in each
community according to their degrees during the anonymization process takes $%
O(n)$ space due to each vertex being involved in only one community. In
addition, since operations Adding Edge and Splitting Vertex create new edges
and vertices during anonymization, to anonymize a vertex $v$, Adding Edge
introduces at most $d_{\max }$ new edges to protect $v$ in a $k$-SDA group
of the largest degree, while Splitting Vertex generates at most $d_{\max }$
substitute vertices given $\left\vert E_{v}\right\vert \leq d_{\max }$.
Consequently, the space complexity of the four heuristics is $O(m+nd_{\max
}) $.

After this, we can determine that the time complexity of each of the four
heuristics is $O(kn^2\log n)$ in the following manner. Firstly, EdgeConnect
achieves the graph anonymization by processing the vertices one-by-one. For
each selected vertex $v$ to be anonymized, the number of redirectable edges
is bounded by the number of new edges incident to $v$, which is at most $%
d_{\max}$. Finding the minimal \textsc{mergence} cost for $v$ involves a
test of all generated anonymous groups, which is bounded by $O(n/k)$.
Computing the minimal \textsc{creation} cost is $O(l\log l)$ since the set $%
U $ consists of $k$ not-yet-anonymized vertices with the largest degrees
from $l$ communities. The adjustment of $v$'s degree and the update of
vertices' order in a community can be achieved within $O(n\log n)$ time. As
such, \textsc{mergence} and \textsc{creation} take $O(n\log n)$ and $%
O(kn\log n)$ time, respectively. The time complexity of anonymizing
$v$ is then $O(d_{\max} +n/k+ l\log l +n\log n+kn\log n)$.
Consequently, since $l < n$, the graph anonymization is achieved in
$O(kn^2\log n)$ time.

Second, with the Vertex Splitting operation, CreateBySplit can also
anonymize a selected vertex $v$ by generating a new anonymous group of a
smaller degree. The discovery of the $k$ vertices with the largest degrees
in different communities costs $O(l\log l)$ time. The splitting of $v$,
including the re-distribution of the incident edges to the two substitute
vertices is upper bounded by $O(d_{\max})$. The update of the vertex order is $%
O(n\log n)$. Therefore, the anonymization process of $v$ takes
$O(kn\log n)$ time. As an extension of EdgeConnect, the complexity
of CreateBySplit is thus $O(kn^2\log n)$.

Third, as with EdgeConnect, MergeBySplit achieves the anonymization of each
selected vertex $v$ in $O(kn\log n)$ time by operation Adding Edge alone. By
Vertex Splitting operation, MergeBySplit anonymizes a selected vertex $v$ in
$O(n\log n)$ time because the minimal number of substitute vertices of $v$
can be determined in $O(n)$, and the re-distribution of incident edges and
the update of vertex order is upper bounded by $O(n\log n)$. Consequently,
the whole graph anonymization is achieved in $O(kn^2\log n)$ time.

Finally, by the operation Adding Edge alone, FlexSplit also
anonymizes each selected vertex $v$ in $O(kn\log n)$ as the
MergeBySplit algorithm. By operation Splitting Vertex, FlexSplit
computes $S^C_v$ in $O(k)$ since there are $k$ vertices in $W$. To
find $X \subseteq W$, it takes $O(kn)$ time to check whether the
vertices in $W$ can be anonymized by \textsc{Creation} and
\textsc{Mergence}, because there are $k$ vertices in $W$ and for
each vertex $u$, it scans $O(n)$ subsequence vertices in the same
community of $u$ to adjust the vertex degree of $u$. After finding
$X$, FlexSplit calculates $\sum_{u \in X} |S^M_u|$ in $O(kn)$ as the
minimal number of substitute vertices of every $u$ in $X$ can be
determined in $O(n)$ according to Formula (\ref{eq:ms}). Thereafter,
FlexSplit chooses between Single Splitting and Group Splitting.
Single Splitting takes $O(n\log n)$ time as in MergeBySplit. Given
$W$, Group Splitting splits the vertices in $W$ in $O(kn)$, and
updates the vertex order in the corresponding communities in
$O(kn\log n)$. Consequently, the overall anonymization time is
bounded by $O(kn^2\log n)$.


\vspace{-2mm}

\section{Experiments}

In this paper, we conduct the experiments on both real and synthetic data
sets. All the social graphs are pre-processed into simple graphs, i.e.
unweighted undirected graphs without self-loops and multiple edges. The
community identities of the vertices are either known as background
knowledge or derived by community detection techniques\footnote{%
METIS graph partition tool, http://glaros.dtc.umn.edu/gkhome/views/metis.}.

\textbf{DBLP:} From the DBLP data set, we select authors who have ever
published their papers in the 20 top conferences, such as AAAI, SIGIR, and
ICDM. The selected data set consists of 30,749 authors, and there are
157,058 edges representing the co-author relationships. As people usually
publish their papers in the conferences related to their interests, we
regard the conference where an author published most of his papers as the
community of the author.

\textbf{ca-CondMat:} This data set shows the scientific collaborations
between authors of papers in the Condense Matter category from January 1993
to April 2003. The graph is available at the SNAP (Stanford Network Analysis
Package) web page, 
and consists of 23,133 vertices and 186,936 edges. An edge is built between
two authors if they had co-authored a paper in that period. Note that the
community (conference) information for this data set is not provided on the
website. We then derive the community identifications by the METIS graph
partition tool, as people in the same social network group or cluster tend
to interact more intensely, i.e., each group or cluster often forms a dense
subgraph.

\textbf{AirPort:} This graph is built by considering the 500 busiest US
airports\footnote{%
http://www.db.cs.cmu.edu/db-site/Datasets/graphData/}. In the graph, there
are 500 vertices representing the airports and 2,980 edges between airports
that have air travel connections. We also derive the community identities by
the METIS graph partition tool.


\textbf{LesMis \cite{for-lm}:} LesMis is a small pseudo social network that
simulates the relationships between 77 characters in Victor Hugo's novel
"Les Miserables." Two characters are linked by an edge if they appear in the
same chapter. There are 254 edges in total. The community information is
derived by the METIS graph partition tool.

In addition, we also use R-MAT graph model \cite{RMAT} to generate synthetic
data sets. R-MAT graph model takes four parameters $a$, $b$, $c$ and $d$,
where $a+b+c+d=1$, to generate graphs that match power-law degree
distributions and small-world properties, observed from many real social
networks. In this paper, we use the default values of 0.45, 0.15, 0.15 and
0.25 for the four corresponding parameters, and generate graphs with the
number of vertices ranging from 20,000 to 100,000 for testing the
scalability of our algorithms.

\vspace{-2mm}

\subsection{Privacy Violation in Real Social Networks}

In this paper, we show that the structural diversity is a real privacy
protection issue against degree attacks in publishing social networks. The
experiments are conducted on two real data sets, DBLP and ca-CondMat.

\begin{figure}[!tbh]
\begin{center}
$%
\begin{array}{cc}
\includegraphics[width=0.235\textwidth]{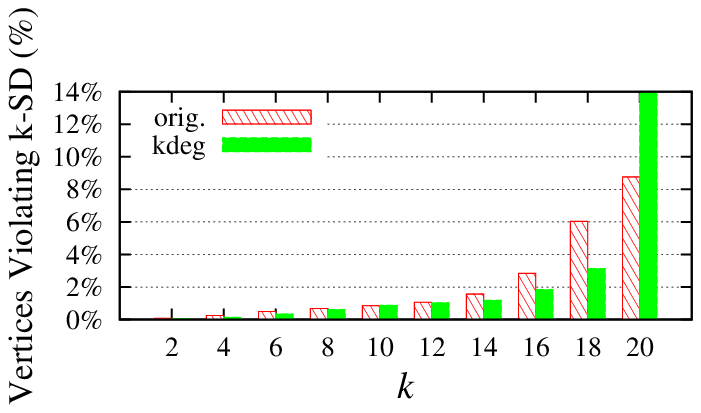} & %
\includegraphics[width=0.235\textwidth]{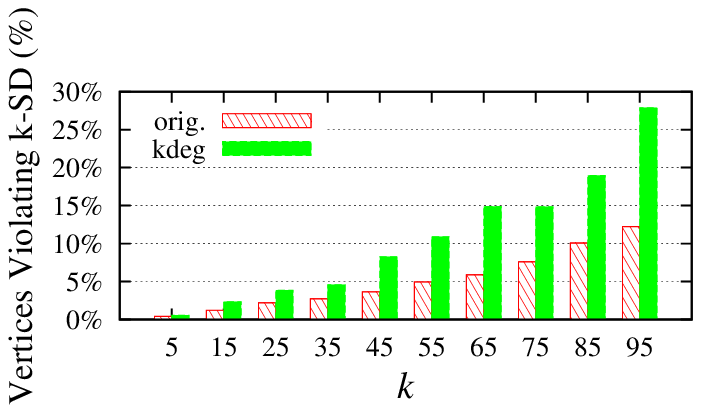} \\
{(a)} & {(b)}%
\end{array}
$%
\end{center}
\par
\caption{Vertices violating $k$ structural diversity on (a) DBLP and
(b) ca-CondMat data sets.} \label{VVSD}
\end{figure}

First, we study the problem of ``whether many vertices of the same degree
tend to gather in the same dense subgraph (community)''. Note that if an
attacker finds all the vertices of a particular degree appearing in a
certain subgraph (community), he can obtain the privacy information such as
the neighborhood and connectivity properties of a target. Privacy will thus
be violated. Figures \ref{VVSD}(a) and \ref{VVSD}(b) show the percentages of
vertices violating $k$-structural diversity ($k$-SD), i.e., the anonymized
group that does not spread over $k$ communities, on the DBLP and ca-CondMat
data sets, respectively. Consider the DBLP data set with $k$ set as 10. In
both the original graph and the 20-degree anonymized graph, there are at
least 2552 (8.3\%) vertices violating 20-SD. As the value of $k$ increases,
the number of vertices violating $k$-SD grows significantly. Figure \ref%
{VVSD} also shows that $k$-degree anonymity sometimes makes this problem
more serious, because $k$-degree anonymity is designed to minimize the
additional edges and does not aim to widely distribute the anonymous
vertices of the same degree. This problem is even more serious for the
ca-Condmat data set.

\begin{figure}[!tbh]
\begin{center}
$%
\begin{array}{cc}
\includegraphics[width=0.235\textwidth]{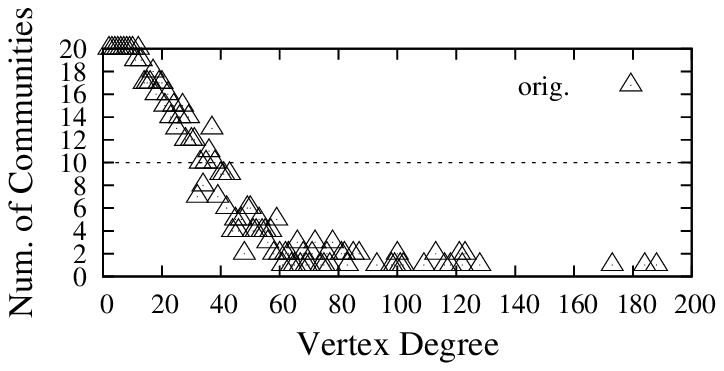} & %
\includegraphics[width=0.235\textwidth]{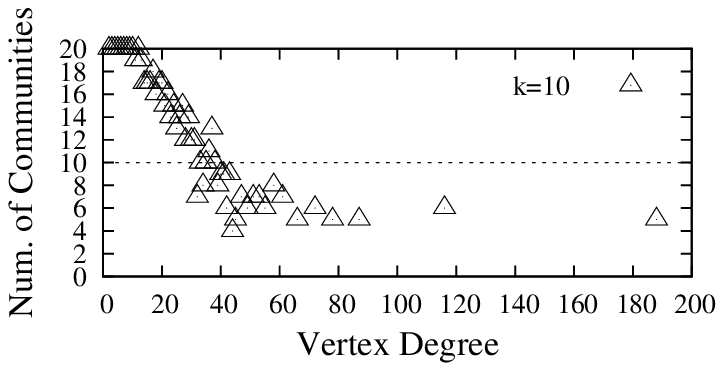} \\
{(a)} & {(b)}%
\end{array}
$%
\end{center}
\par
\caption{Vertices with the same degree over the number of
communities in (a) original DBLP and (b) DBLP protected by 10-degree
anonymity.} \label{VVSD2}
\end{figure}

Next, we study the problem of ``what the degrees are of the vertices
violating $k$-SD''. In this experiment, we test the DBLP data set without
anonymization and with 10-degree anonymization. Figure \ref{VVSD2} shows the
number of communities containing vertices of a particular degree. Consider
the case of 10-SD. The data points with the community numbers smaller than
10 (below the horizontal dashed line) violate 10-SD. It is worth mentioning
that the vertices violating 10-SD have large degrees. This means that active
people are more likely to have higher risks of privacy violation.


In summary, the experimental results show that the structural diversity is a
real privacy protection issue against degree attacks, especially for the
vertices of large degrees. Moreover, graphs protected by $k$-degree
anonymity may still violate $k$-SD as $k$-degree anonymity is not designed
for the $k$-SDA problem.



\subsection{Anonymization Performance}

In this subsection, we evaluate the performance of the EdgeConnect (EC),
CreateBySplit (CBS), MergeBySplit (MBS) and FlexSplit (FS) algorithms
compared with the optimal solution, $k$-degree anonymity\footnote{%
We implement the Priority algorithm in \cite{k-deg}.}, Algorithm Inverse
EdgeConnect (IEC)\footnote{%
EC increases the degree of a vertex $v$ from $d_v$ to $\widehat{d}$ by
connecting $v$ with not-yet-anonymized vertices of the largest degrees,
while IEC connects $v$ and the last ($\widehat{d}-d_{v}$) vertices in the
sequence with the smallest degrees.} and SplittingOnly (Sonly)\footnote{%
Sonly extracts only the capabilities of the flexible splitting strategy in
FlexSplit and does not apply operation Adding Edge.}. 



\subsubsection{Utility Studies}

We now study the utility of anonymized graphs from the clustering
coefficients (CC), average shortest path lengths between vertex pairs
(ASPL), betweenness centralities (BC), degree centralities (DC), eigenvector
centrality correlations with respect to original graphs (EC-correlation),
degree frequencies, the accuracy of community detection and connected query
results on the DBLP and ca-CondMat data sets. In all of the above
evaluations, we also compare our four heuristic algorithms with $k$-degree
anonymity.

\begin{figure*}[tbh]
\begin{center}
$%
\begin{array}{c}
\includegraphics[width=0.2\textwidth]{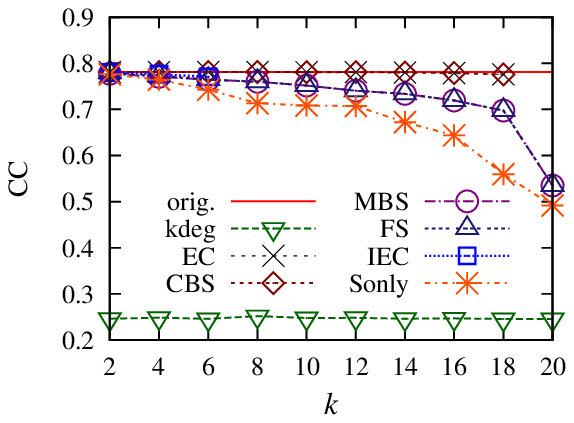} %
\includegraphics[width=0.2\textwidth]{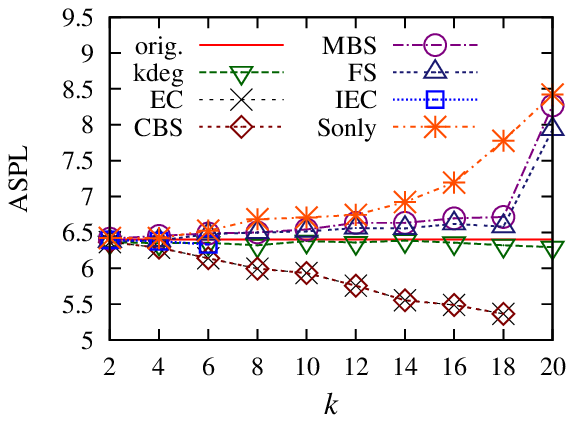} %
\includegraphics[width=0.2\textwidth]{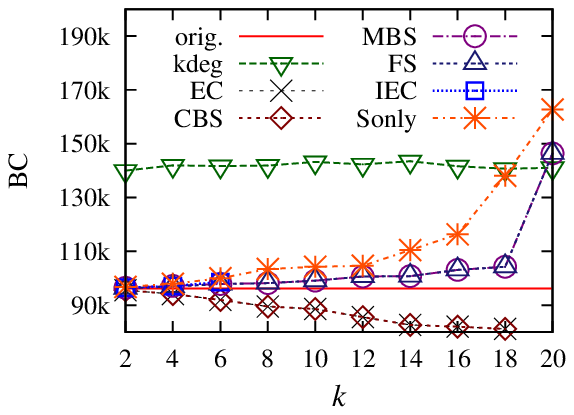} %
\includegraphics[width=0.2\textwidth]{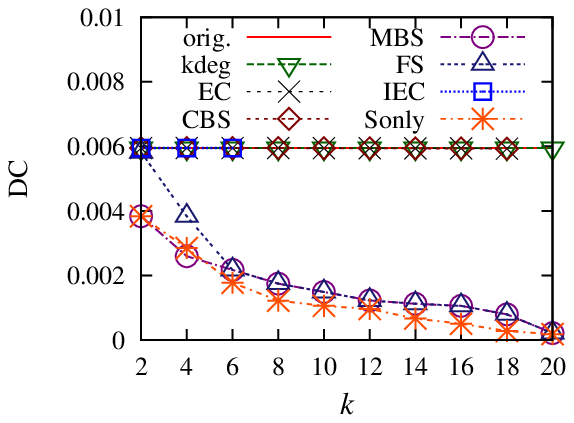} %
\includegraphics[width=0.2\textwidth]{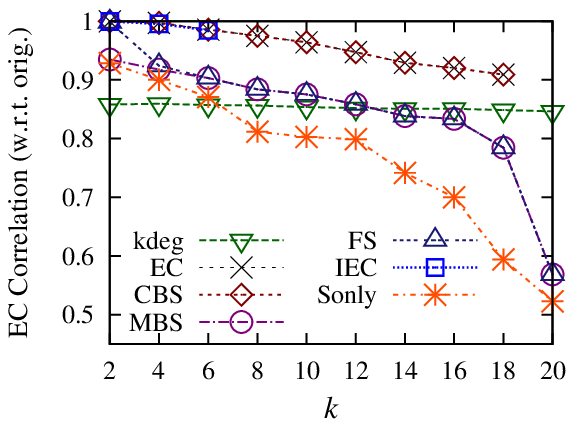} \\
{(a)\hspace{30mm}(b)\hspace{30mm}(c)\hspace{30mm}(d)\hspace{30mm}(e)} \\
\includegraphics[width=0.2\textwidth]{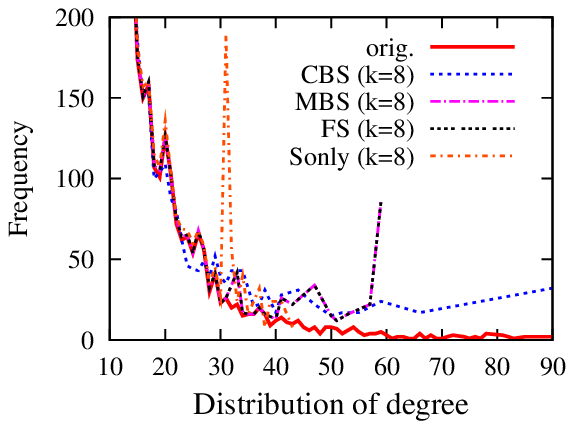} %
\includegraphics[width=0.2\textwidth]{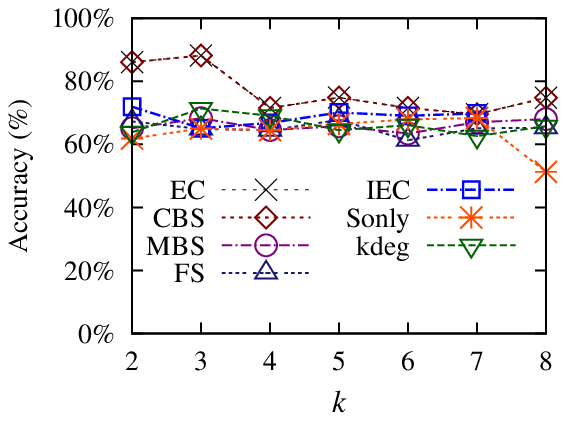} %
\includegraphics[width=0.2\textwidth]{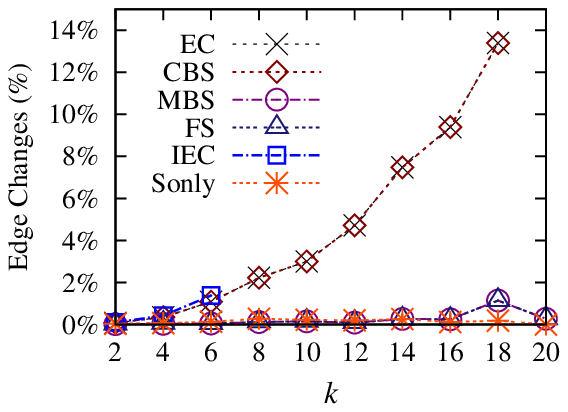} %
\includegraphics[width=0.2\textwidth]{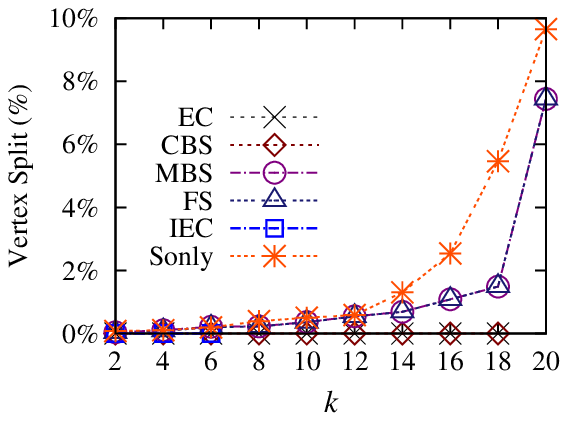} %
\includegraphics[width=0.2\textwidth]{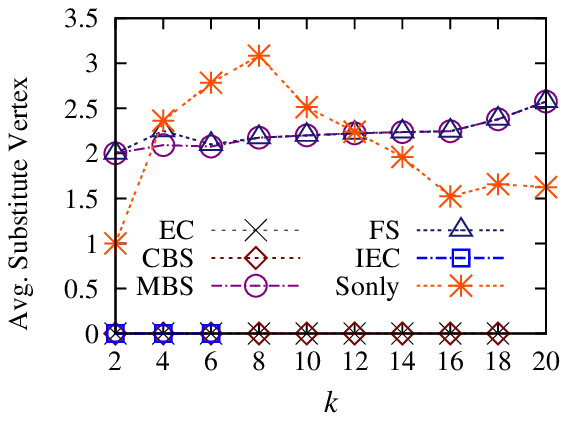} \\
{(f)\hspace{30mm}(g)\hspace{30mm}(h)\hspace{30mm}(i)\hspace{30mm}(j)}%
\end{array}
$%
\end{center}
\par
\vspace{-4mm} \caption{Performance evaluations on DBLP. }
\label{dblp}
\end{figure*}
\begin{figure*}[tbh]
\begin{center}
$%
\begin{array}{c}
\includegraphics[width=0.2\textwidth]{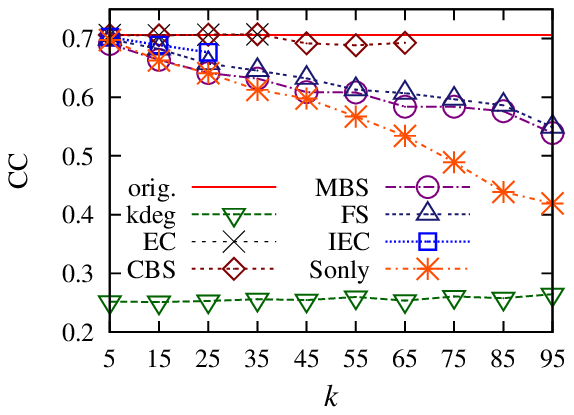} %
\includegraphics[width=0.2\textwidth]{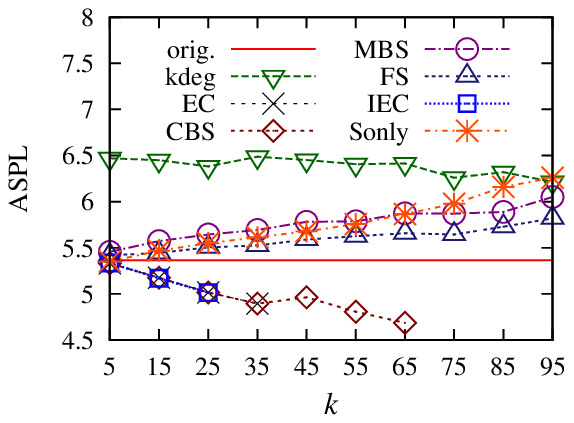} %
\includegraphics[width=0.2\textwidth]{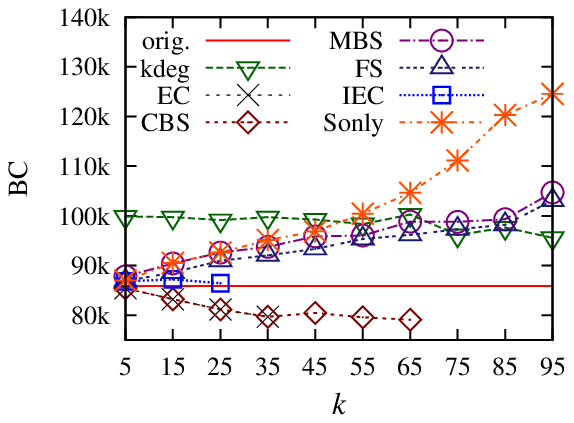} %
\includegraphics[width=0.2\textwidth]{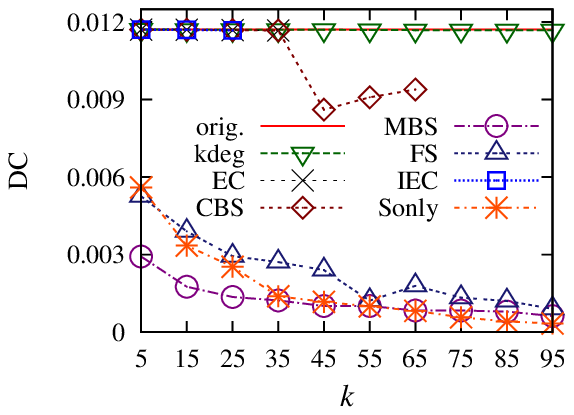} %
\includegraphics[width=0.2\textwidth]{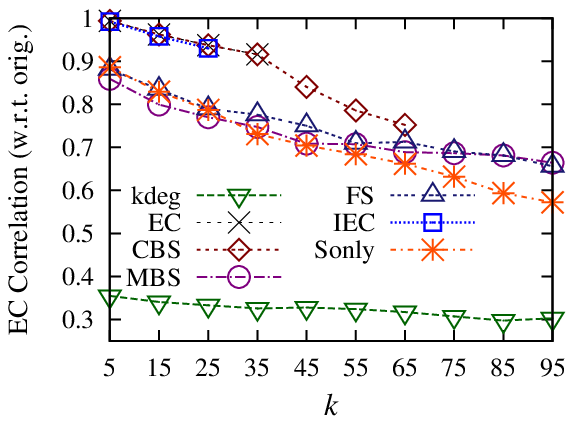} \\
{(a)\hspace{30mm}(b)\hspace{30mm}(c)\hspace{30mm}(d)\hspace{30mm}(e)} \\
\includegraphics[width=0.2\textwidth]{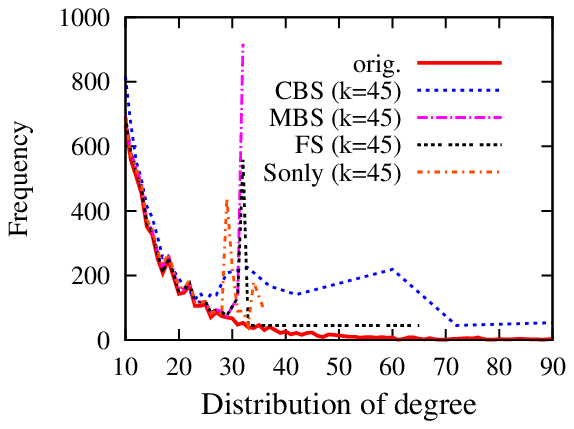} %
\includegraphics[width=0.2\textwidth]{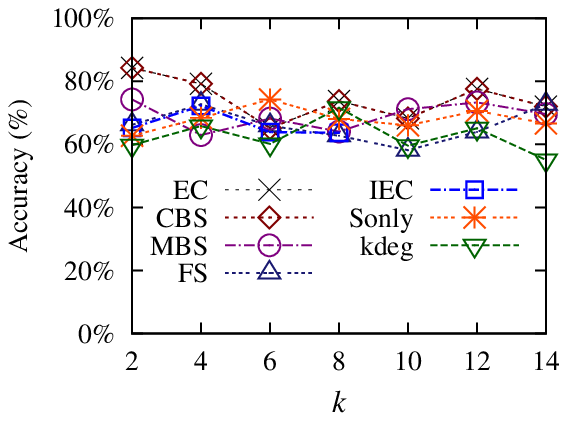} %
\includegraphics[width=0.2\textwidth]{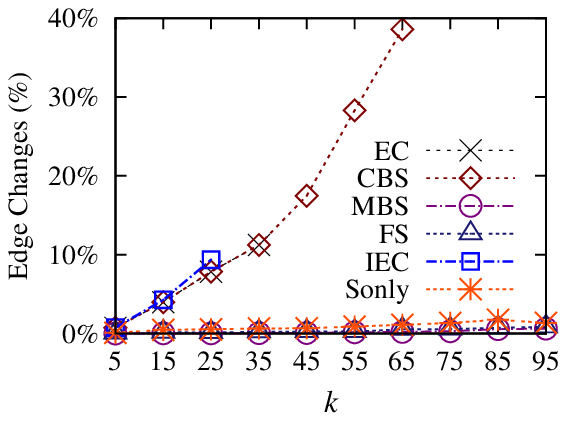} %
\includegraphics[width=0.2\textwidth]{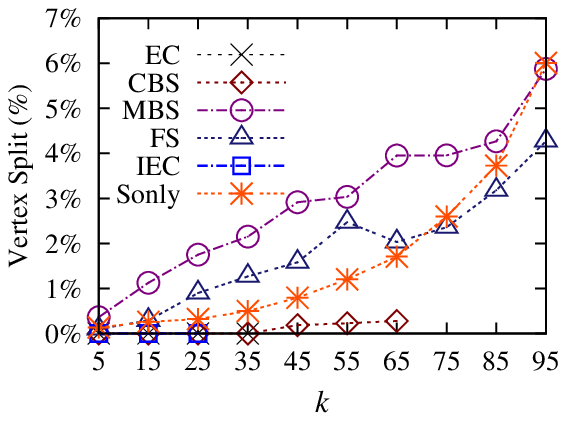} %
\includegraphics[width=0.2\textwidth]{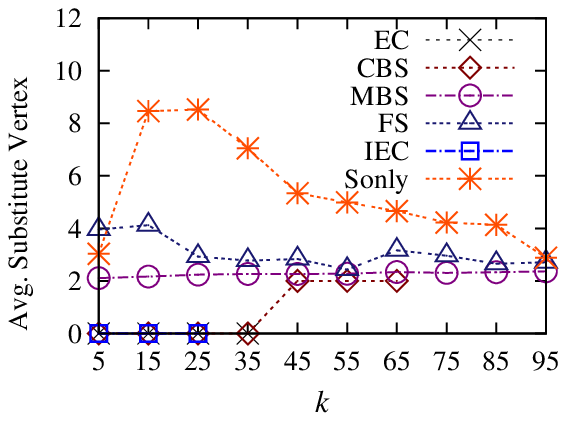} \\
{(f)\hspace{30mm}(g)\hspace{30mm}(h)\hspace{30mm}(i)\hspace{30mm}(j)}%
\end{array}
$%
\end{center}
\par
\vspace{-4mm} \caption{Performance evaluations on ca-CondMat. }
\label{ca}
\end{figure*}

\textbf{Clustering Coefficient (CC):} Figures \ref{dblp}(a) and \ref{ca}(a)
show the clustering coefficients of the anonymized DBLP and ca-CondMat as a
function of $k$, respectively. The CC values of the original DBLP and
ca-CondMat are about 0.781 and 0.706. It should be first pointed out that EC
can almost perfectly preserve the clustering coefficient of the original
graphs on both data sets. This is because EC only adds new edges within
communities for anonymization and thus preserves many of the community
structures. The trade-off, however, is that on ca-CondMat, EC anonymizes the
graph successfully only when $k$ is (relatively) small. As an extension, CBS
has a greater chance to achieve the anonymization when $k$ becomes larger,
as is evident from Figure \ref{ca}(a), while the cost is a small decrease in
the CC values due to the splitting of some vertices. To guarantee the
anonymization, MBS does not connect the substitute vertices of each split
vertex and, therefore, weakens the cohesiveness of the communities
especially when $k$ grows closer to the total number of communities in the
graphs. Compared to MBS, FS has the CC values closer to the original value
as FS reduces the numbers of substitute vertices in the objective function
of $k$-SDA.
Finally, note that our four algorithms all outperform $k$-degree anonymity
in preserving the community structures.

\textbf{Average Shortest Path Lengths (ASPL):} Figures \ref{dblp}(b) and \ref%
{ca}(b) show the average shortest path lengths between vertex pairs of the
anonymized DBLP and ca-CondMat as a function of $k$, respectively. The ASPLs
of the original DBLP and ca-CondMat are about 6.4 and 5.36. EC monotonically
decreases the ASPL values as $k$ grows because edges within communities are
added for anonymization. CBS has better EC while the ASPL values neither
monotonically decrease nor increase. This is because CBS not only introduces
new edges within communities but also splits vertices and connects the
substitute vertices of each split vertex. The cost of MBS for the guarantee
of anonymization is the increase of the ASPL values, as the substitute
vertices do not directly connect to each other.
By reducing the numbers of substitute vertices, FS has the ASPL values
closer to those of the original graph than those of MBS. Finally, $k$-degree
anonymity performs quite well on the DBLP data set, as depicted in Figure %
\ref{dblp}(b), because $k$-degree anonymity provides less protection and
requires only a few additional edges for anonymization. On ca-CondMat,
however, the proposed methods all perform better than $k$-degree anonymity.
The reason for this is that we consider the community structures and connect
only the vertices in the neighborhoods.

\textbf{Betweenness Centrality (BC):} Figures \ref{dblp}(c) and \ref{ca}(c)
show the betweenness centralities, i.e., the frequency of a vertex on the
shortest paths between pairs of vertices, of the anonymized DBLP and
ca-CondMat as a function of $k$, respectively. For similar reasons mentioned
in the ASPL measurement, here we observe that the BC values of the four
proposed algorithms have similar trends (with respect to the original value)
as the ASPL values, and the proposed methods preserve BC better than the $k$%
-degree method.

\textbf{Degree Centrality (DC):} For a graph, a large degree centrality,
which is usually used to measure the influential vertices in social network
analysis, indicates the existence of vertices with relatively large degrees.
The DC comparisons of the anonymized DBLP and ca-CondMat obtained by the
proposed four methods and $k$-degree anonymization are presented in Figures %
\ref{dblp}(d) and \ref{ca}(d). The original DC values of DBLP and ca-CondMat
are 0.00594 and 0.011713, respectively. On both data sets, EC, CBS and $k$%
-degree anonymization perform perfectly. This indicates that the three
methods can effectively preserve the strong leaders and influential vertices
in the social networks. In contrast, MBS and FS sacrifice the precision of
DC in order to guarantee the anonymization. In other words, anonymizing the
vertices in increasing order of the degrees tends to make the vertices have
similar small degrees by the Splitting Vertex operation. Nonetheless, FS
still outperforms MBS for many cases.


\textbf{Eigenvector Centrality Correlation (EC-Correlation):} Eigenvector
centrality, another common measurement of influential vertices in the social
networks, estimates the influence of a vertex based on the influence of the
vertices to which the directed neighbors connect. Figures \ref{dblp}(e) and %
\ref{ca}(e) show the EC-correlations of the anonymized DBLP and ca-CondMat
(with respect to the original graph). It can be seen that EC has the
EC-correlations above 0.9 and achieves the best preservation of influential
vertices. The other three methods have the EC-correlations above 0.7 for
most cases. Differing from the results in the DC measurement, here the four
proposed methods all outperform the $k$-degree anonymity, as a result of the
structural information being taken into account in the anonymization.


\textbf{Degree Frequency (DF):} Figures \ref{dblp}(f) and \ref{ca}(f)
compare the degree distributions of anonymized DBLP and ca-CondMat with the
original graph, respectively. Although the distributions in small degrees
are similar to the original distributions, due to the different splitting
strategies, CBS performs better than FS, and FS outperforms MBS in
preserving the distributions in large degrees.

\textbf{Community Detection:} Figures \ref{dblp}(g), \ref{ca}(g), \ref{ap}%
(g) and \ref{lm}(g) present the accuracy of community detection on the
anonymized graphs with respect to the original DBLP, ca-CondMat, AirPort,
and LesMis graphs, respectively. The results indicate that all heuristics
achieve comparable performance to the optimal solution (in Figures \ref{ap}%
(g) and \ref{lm}(g)) and $k$-degree anonymity (in Figures \ref{dblp}(g) and %
\ref{ca}(g)), while the heuristics are able to provide stronger privacy
protection than $k$-degree anonymity. EC always outperforms $k$-degree
anonymity on maintaining the community structures, demonstrating that adding
edges within a community can preserve semantic meanings. More interestingly,
EC slightly outperforms the optimal solution on AirPort in Figure \ref{ap}%
(g). This may indicate that, in addition to the number of new edges
involved, the selection of the vertices to be connected and the vertices to
be split is also crucial for preserving communities in anonymized graphs. In
Figure \ref{ca}(g), the heuristics still outperform $k$-degree anonymity in
most cases when Splitting Vertex is incorporated. When a vertex can be
split, the accuracy of all heuristics is not lowered as $k$ increases.


\textbf{Connected Query:} In addition to the measurements above, it is worth
specifically mentioning that FS also outperforms MBS in the capability of
answering queries for pairs of vertices. For the ca-CondMat data set, about
0.01\% to 0.03\% (among two hundred million) pairs of connected vertices
will be disconnected in the anonymization process of MBS when $k$ varies
from 5 to 95, while none is disconnected by FS. This is because
MBS does not directly link the substitute vertices, and FS
is able to reduce the numbers of substitute vertices with Group Splitting,
which connects substitute vertices.



In light of the above evaluations, CreateBySplit outperforms EdgeConnect in
guaranteeing the anonymization, while FlexSplit can preserve the utility of
a social network better than MergeBySplit. Therefore, we recommend
CreateBySplit for the cases of (relatively) small $k$ and FlexSplit for more
challenging cases.

\subsubsection{Vertex Change and Edge Change}

We now report on the three findings of (a) the percentage of the number of
new edges to the original number of edges, (b) the percentage of the number
of vertices being split to the original number of vertices, and (c) the
average number of substitute vertices for a vertex split, of the anonymized
graphs as functions of $k$.

For DBLP, first, Figure \ref{dblp}(g) shows that when the value of $k$ is
smaller than 50\% of the number of communities, EC and CBS achieve the $k$%
-structural diversity by adding less than 5\% new edges in the anonymized
graph. Second, the results in Figures \ref{dblp}(g) and \ref{dblp}(h) show
that when $k$ becomes larger, CBS tends to add new edges rather than to
split the vertices, while MBS and FS are prone to splitting vertices rather
than to adding new edges. This difference in tendency is caused by the
reverse order of creating the anonymous groups of particular degrees, as we
have more chances to add new edges for the anonymization when the vertices
are anonymized in the decreasing order of the degrees. Third, Figures \ref%
{dblp}(h) and \ref{dblp}(i) show that MBS and FS use a similar number of
substitute vertices for a similar percentage of vertices that have been
split. On DBLP, MBS and FS thus achieve comparable performances for most $k$.

For ca-CondMat, the four algorithms have similar trends of adding edges and
splitting vertices as those for DBLP. However, Figure \ref{ca}(h) shows that
FS splits 1\% to 2\% fewer vertices than MBS under the same guarantee of
anonymization. 
Moreover, in Figure \ref{ca}(i), FS uses more substitute vertices on average
for a vertex that has been split. This indicates that a vertex being split
is likely to be a vertex with a large degree. The results also conform to
those described in Figure \ref{ca}(f).

\subsubsection{Comparison with Optimal Solution}

Here we compare the heuristics with the Integer Programming method, while
the optimal solution is obtained with the proposed formulation using CPLEX%
\footnote{%
http://www-01.ibm.com/software/integration/optimization/cplex/.}. Note that
finding the optimal solutions is very computationally intensive (e.g., for
the AirPort dataset consisting of 500 vertices and 2,980 edges, it takes at
least one hour for the simplest instance and at least one day for more
challenging instances). The optimal solutions are not able to be returned
within a reasonable time frame for large social networks, such as DBLP and
ca-CondMat. Therefore, the solutions from the proposed algorithms are
compared with the optimal solutions of AirPort and LesMis, with $k$ from 2
to 4.

\begin{figure*}[tbh]
\begin{center}
$%
\begin{array}{c}
\includegraphics[width=0.2\textwidth]{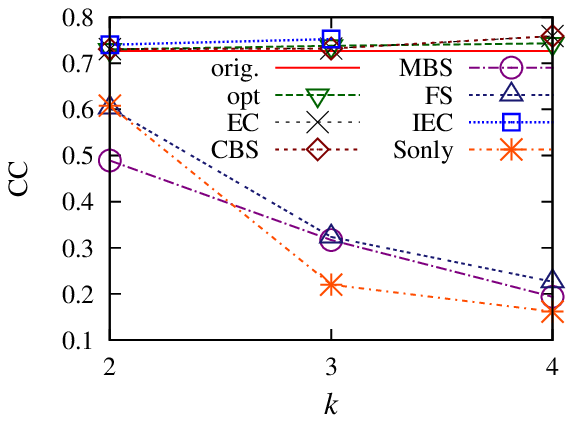} %
\includegraphics[width=0.2\textwidth]{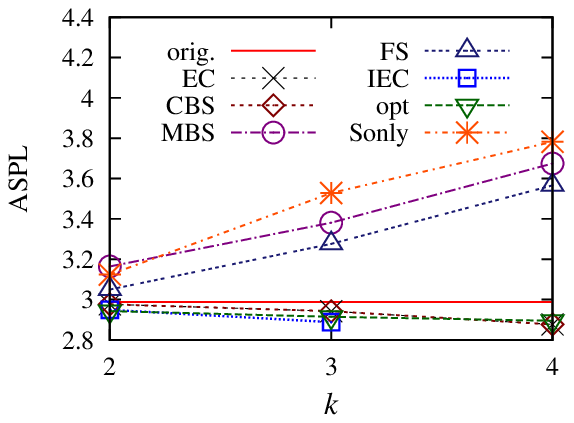} %
\includegraphics[width=0.2\textwidth]{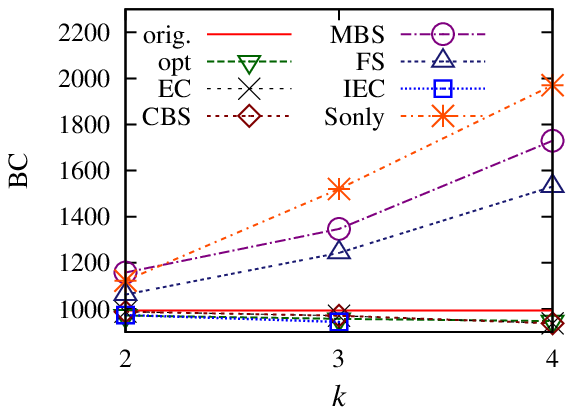} %
\includegraphics[width=0.2\textwidth]{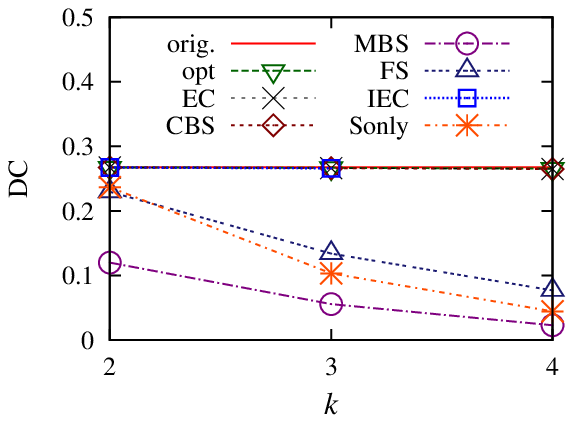} %
\includegraphics[width=0.2\textwidth]{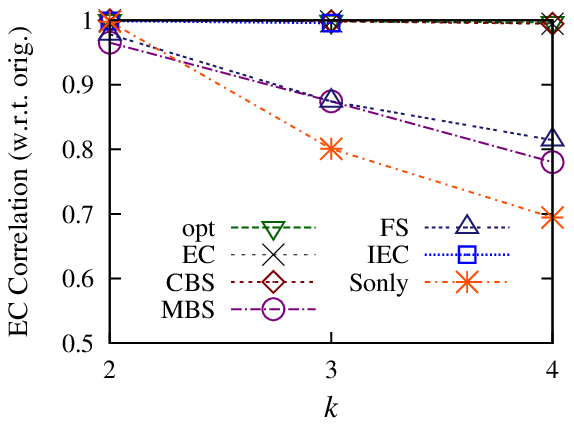} \\
{(a)\hspace{30mm}(b)\hspace{30mm}(c)\hspace{30mm}(d)\hspace{30mm}(e)} \\
\includegraphics[width=0.2\textwidth]{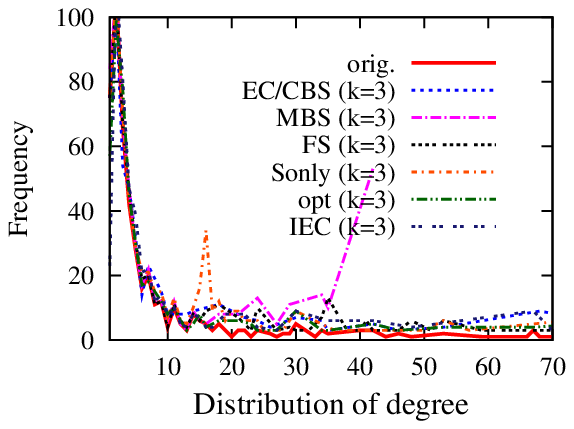} %
\includegraphics[width=0.2\textwidth]{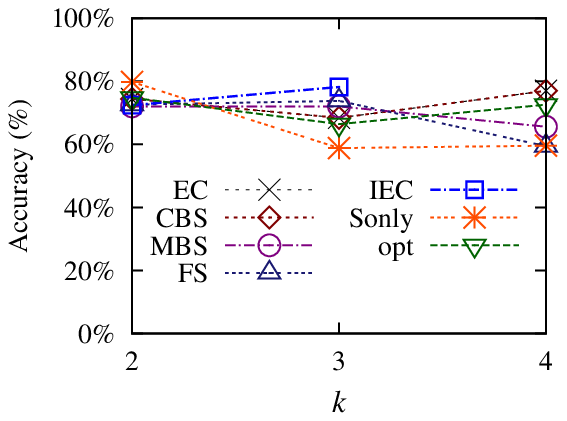} %
\includegraphics[width=0.2\textwidth]{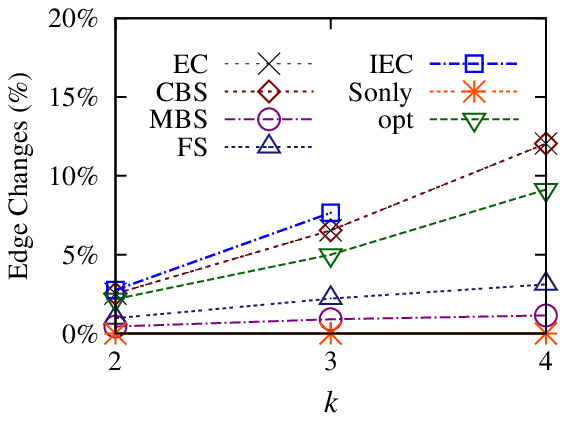} %
\includegraphics[width=0.2\textwidth]{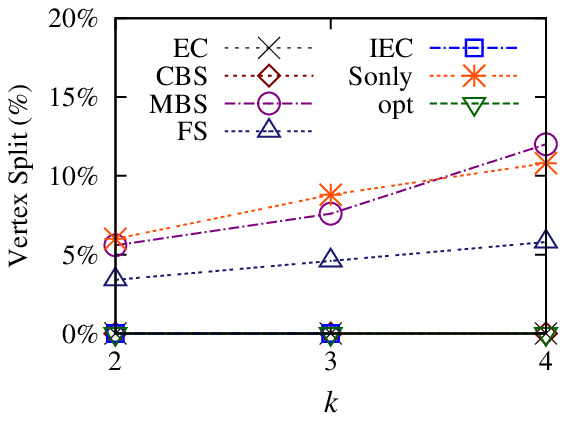} %
\includegraphics[width=0.2\textwidth]{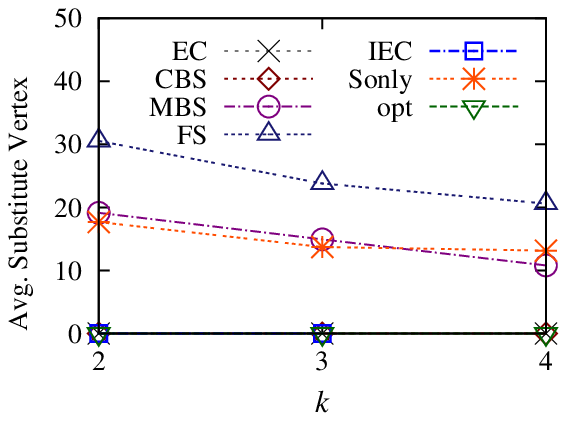} \\
{(f)\hspace{30mm}(g)\hspace{30mm}(h)\hspace{30mm}(i)\hspace{30mm}(j)} \\
\end{array}
$%
\end{center}
\par
\vspace{-4mm} \caption{Performance evaluations on AirPort. }
\label{ap}
\end{figure*}

\begin{figure*}[tbh]
\begin{center}
$%
\begin{array}{c}
\includegraphics[width=0.2\textwidth]{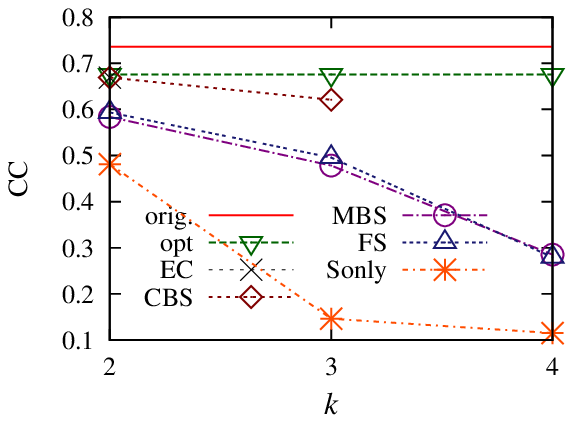} %
\includegraphics[width=0.2\textwidth]{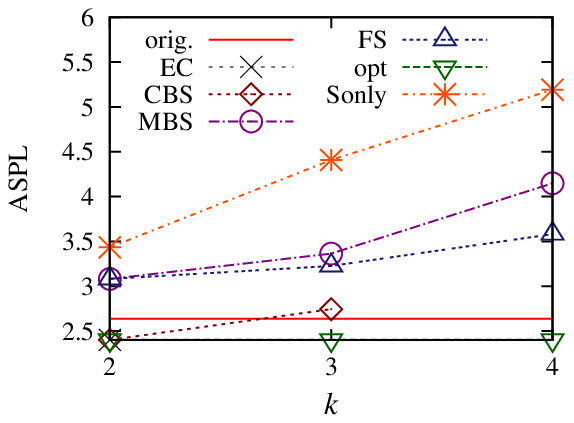} %
\includegraphics[width=0.2\textwidth]{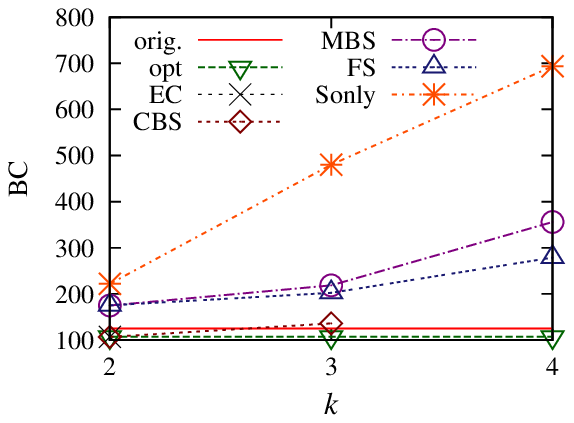} %
\includegraphics[width=0.2\textwidth]{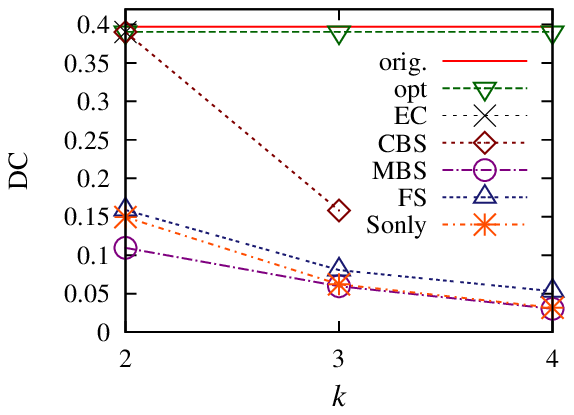} %
\includegraphics[width=0.2\textwidth]{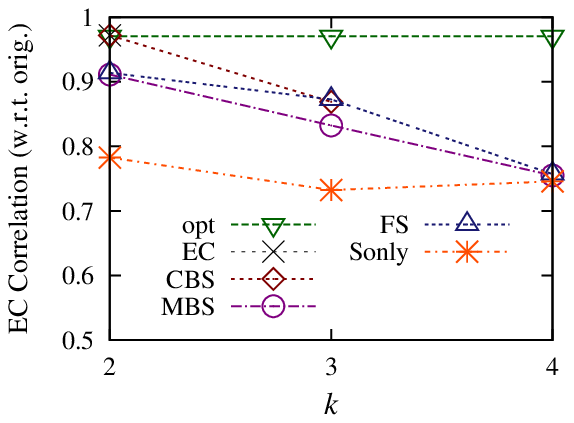} \\
{(a)\hspace{30mm}(b)\hspace{30mm}(c)\hspace{30mm}(d)\hspace{30mm}(e)} \\
\includegraphics[width=0.2\textwidth]{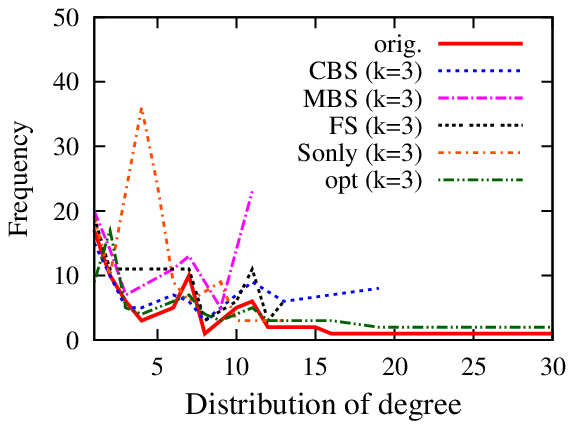} %
\includegraphics[width=0.2\textwidth]{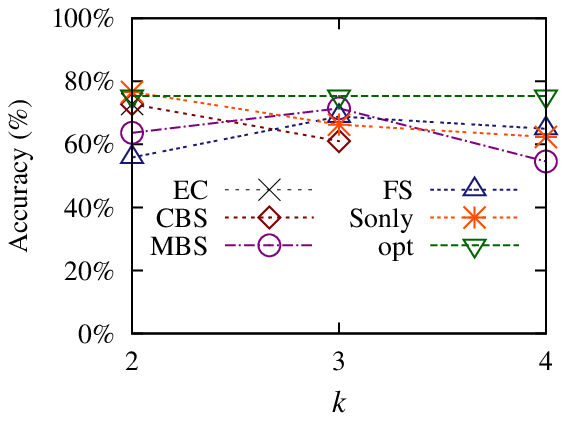} %
\includegraphics[width=0.2\textwidth]{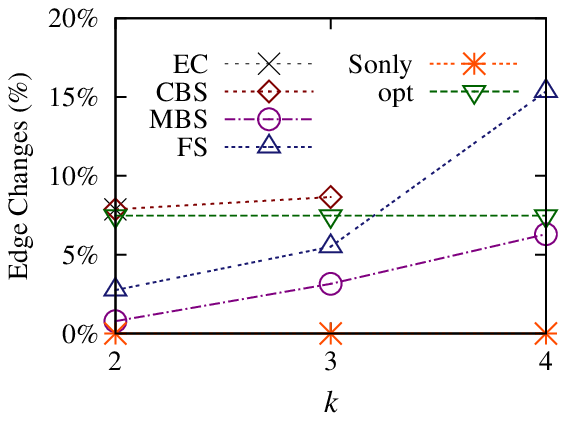} %
\includegraphics[width=0.2\textwidth]{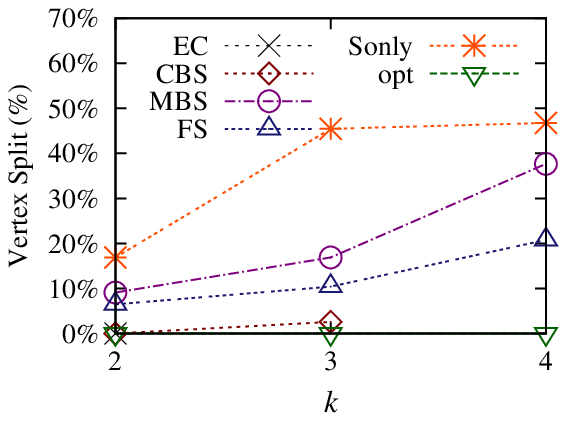} %
\includegraphics[width=0.2\textwidth]{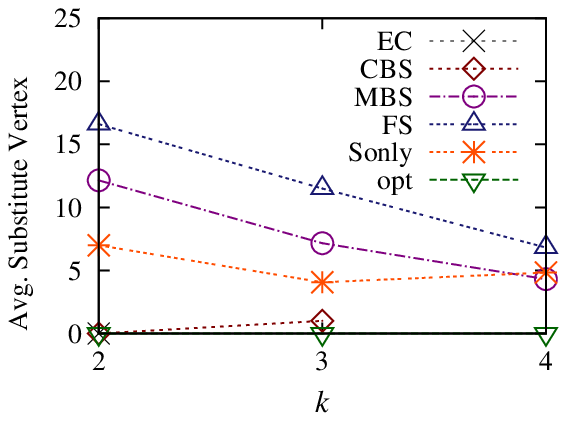} \\
{(f)\hspace{30mm}(g)\hspace{30mm}(h)\hspace{30mm}(i)\hspace{30mm}(j)} \\
\end{array}
$%
\end{center}
\par
\vspace{-4mm} \caption{Performance evaluations on LesMis. }
\label{lm}
\end{figure*}

Figures \ref{ap}(a)-\ref{ap}(g) and \ref{lm}(a)-\ref{lm}(g) respectively
present the data utility of the anonymized graphs of AirPort and LesMis in
terms of the clustering coefficients (CC), average shortest path lengths
between vertex pairs (ASPL), betweenness centralities (BC), degree
centralities (DC), eigenvector centrality correlations with respect to the
original graphs (EC-correlation), degree frequency distributions, and
community detection accuracy. It can be observed that EC is close to the
optimal solution in all evaluations but fails to anonymize LesMis when $k$
is set as 3 and 4, because EC applies only Adding Edge with edge-redirection
to reduce the number of new edges. Moreover, for AirPort with all $k$ and
LesMis with $k$ = 2, CBS is very close to the optimal solution because CBS
applies Splitting Vertex only when Adding Edge alone cannot achieve the
anonymization. In contrast, MBS and FS deviate from the optimal solutions,
because these heuristics apply Splitting Vertex and begin the anonymization
from vertices of small degrees in order to guarantee the success of
anonymization for any instance. Here the results are consistent with those
obtained on the large social networks of DBLP and ca-CondMat.


\subsubsection{Comparison of EC with IEC and Sonly}


We compare EdgeConnect (EC) with Inverse EdgeConnect (IEC) and SplittingOnly
(Sonly) to explore the intuition beyond the design of Algorithm EdgeConnect
and the extensions.

First, EC is compared with IEC on DBLP in Figure \ref{dblp}, ca-CondMat in
Figure \ref{ca}, and AirPort in Figure \ref{ap}\footnote{%
The comparison is not performed on LesMis, because IEC is not able to return
feasible solutions on LesMis.}. Indeed, the results indicate that IEC
outperforms EC in terms of the average shortest path length (ASPL) and
betweenness centrality (BC) (for the cases IEC returns a feasible solution,
i.e., when $k$ = 2, 4, 6 in Figure \ref{dblp}, $k$ = 5, 15, 25 in Figure \ref%
{ca}, and $k$ = 2, 3 in Figure \ref{ap}), because EC takes as its priority
choosing the vertices with large degrees. As those vertices are more
inclined to participate in the shortest paths of any two vertices, EC
reduces ASPL and BC in the anonymized graph. Therefore, IEC is suitable for
the application scenarios in which the characteristics of shortest paths are
the major properties required to be preserved during anonymization.

\begin{figure}[h]
\centering
\includegraphics[width=0.4\textwidth]{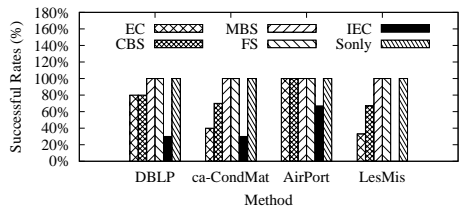}
\caption{Successful rates of heuristics.} \label{srate}
\end{figure}

On the other hand, the clustering coefficient (CC) of the anonymized graph
from EC is closer to the CC value of the original graph, and EC is able to
achieve better accuracy in community detection for most cases, as
demonstrated in Figures \ref{dblp}(g), \ref{ca}(g) and \ref{ap}(g). It is
noteworthy that EC incurs fewer new edges than IEC in Figures \ref{dblp}(h), %
\ref{ca}(h) and \ref{ap}(h), and generates a higher successful rate in
anonymization, as seen in Figure \ref{srate}. The reason is that a new edge
involved in the edge-redirection operation of EC has more opportunities to
be reused in the anonymization of other vertices considered later, as EC
adds new edges between anonymizing vertex $v$ and vertices of large degrees
prior to being anonymized. EC is thus more capable of handling the input
instances that are difficult to be anonymized by introducing only new edges.

The comparisons of EC and Sonly being conducted on DBLP are presented in
Figure \ref{dblp}, on ca-CondMat in Figure \ref{ca}, on AirPort in Figure %
\ref{ap}, and on LesMis in Figure \ref{lm}. Whereas Sonly preserves ASPL and
BC better than EC in DBLP and ca-CondMat when $k$ is small, for AirPort and
LesMis, EC significantly outperforms Sonly. This is because the degree
differences between the vertices of the largest degree and the other
vertices are more significant in DBLP and ca-CondMat datasets, and EC is
prone to connecting the vertices of large degrees to the others, which
thereby significantly shortens many of the shortest paths among the
vertices. %
In contrast, for a small $k$, Sonly only needs to split a few
vertices of the largest degree to fulfill $k$-structural diversity.
This results in the lengths of the shortest paths increasing
slightly. For the other parameters, such as CC, DC, EC-correlation,
degree frequency distribution, and community detection in most
cases, the findings indicate that EC outperforms Sonly because
Splitting Vertex not only decreases the vertex degrees but also
tends to change the community structure. Nevertheless, as
demonstrated in Figure \ref{srate}, operation Splitting Vertex is
necessary in our algorithm design for the social graphs that are
difficult to anonymize.

\subsubsection{Anonymization Successful Rate}


Here we compare the successful rates of the heuristics on DBLP, ca-CondMat,
AirPort and LesMis datasets. The results in Figure \ref{srate} show that
MBS, FS, and Sonly are guaranteed to anonymize any social graph thanks to
operation Splitting Vertex. Those approaches begin the anonymization process
from the vertices of small degrees to generate anonymous groups. For this
reason, the anonymous group of degree 1 will be generated first, and
Splitting Vertex can thus partition a vertex of any degree into multiple
substitute vertices of degree 1, even in the most challenging case in
anonymization. In contrast, EC, CBS and IEC may not always be able to
anonymize a graph. The vertices of large degrees usually appear in the same
community (e.g., a clique), and not every community contains sufficient
vertices of small degrees for anonymization. Therefore, when anonymous
groups of large degrees are generated prior to those of small degrees, the
added new edges within a community may significantly increase the degrees of
the not-yet-anonymized vertices originally with small degrees, such that it
becomes difficult afterward to anonymize other vertices with small degrees.
Compared with the other schemes, the successful rate of IEC is smaller
because IEC takes priority to add new edges connecting to the vertices with
small degrees, thereby further increasing the difficulty to anonymize those
vertices.

\begin{figure*}[tbh]
\begin{center}
$%
\begin{array}{cccc}
\includegraphics[width=0.2\textwidth]{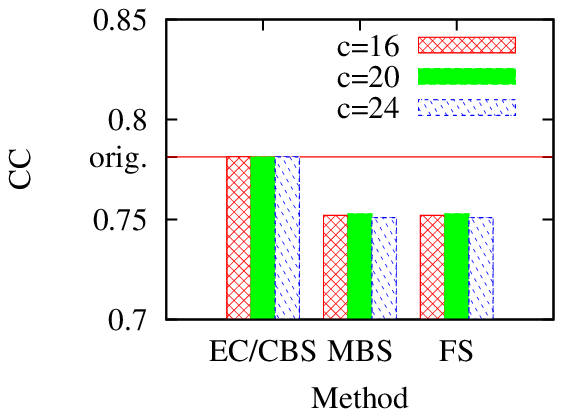} & %
\includegraphics[width=0.2\textwidth]{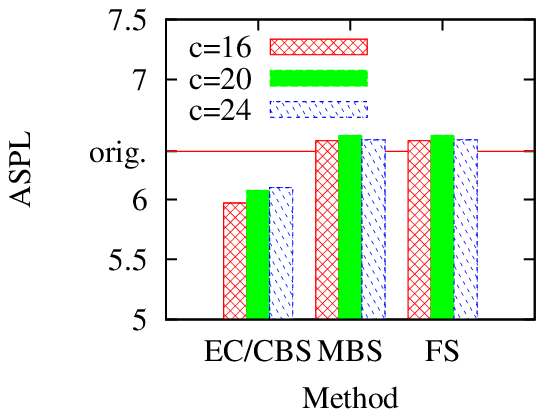} & %
\includegraphics[width=0.2\textwidth]{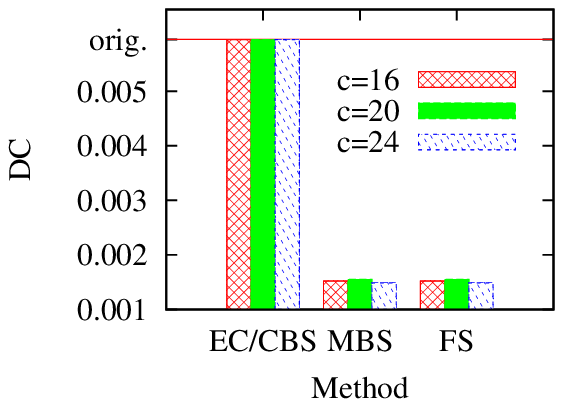} & %
\includegraphics[width=0.2\textwidth]{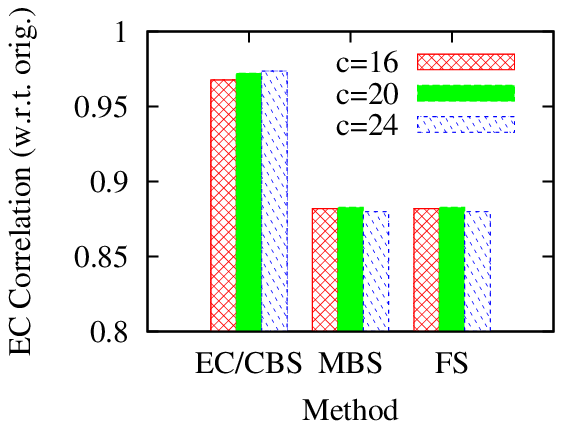} \\
{(a)} & {(b)} & {(c)} & {(d)}%
\end{array}
$%
\end{center}
\par
\vspace{-2mm} \caption{Sensitivity studies given different numbers
of communities.} \label{sensi}
\end{figure*} %
\begin{figure*}[!tbh]
\begin{center}
$%
\begin{array}{cccc}
\includegraphics[width=0.2\textwidth]{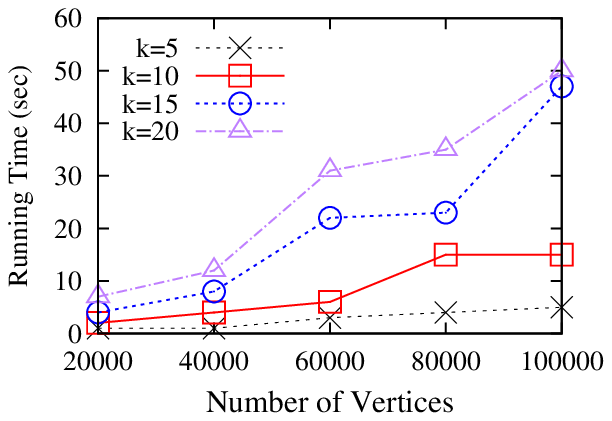} & %
\includegraphics[width=0.2\textwidth]{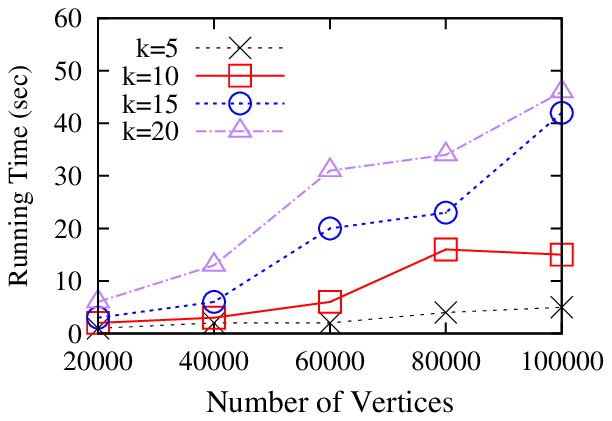} & %
\includegraphics[width=0.2\textwidth]{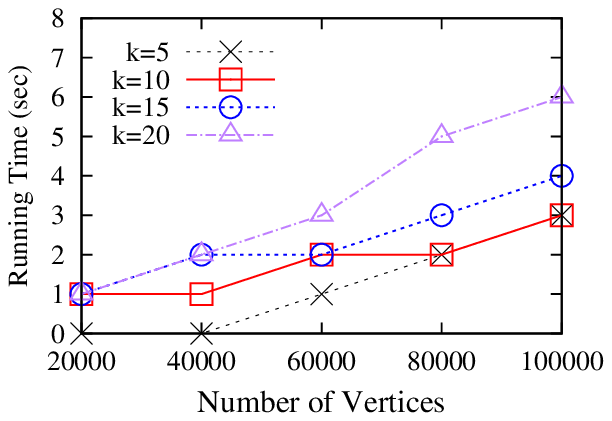} & %
\includegraphics[width=0.2\textwidth]{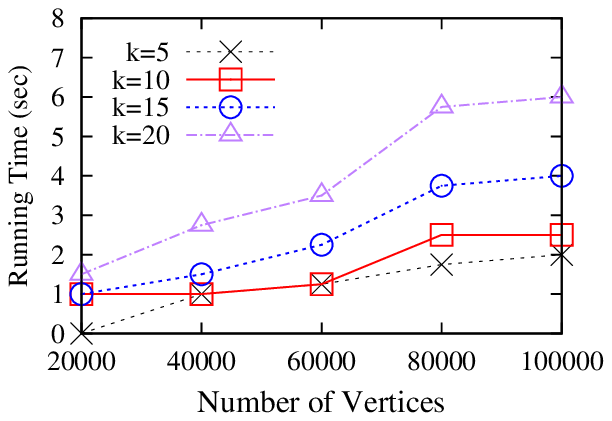} \\
{(a)} & {(b)} & {(c)} & {(d)}%
\end{array}
$%
\end{center}
\par
\vspace{-4mm} \caption{Scalability of Algorithm (a) EdgeConnect, (b)
CreateBySplit, (c) MergeBySplit and (d) FlexSplit.} \label{SCL}
\end{figure*} %

\subsubsection{Sensitivity}

Consider the case that the communities are not given explicitly and,
instead, community detection techniques are used to obtain the community
information for structural diversity. We then explore the sensitivities of
Algorithm EC, CBS, MBS and FS to the number of communities obtained by
community detection techniques. In these experiments, we conduct the
analysis on DBLP as we know the ground truth of the communities in the data
set. Figure \ref{sensi} presents the CC, ASPL, DC and EC-correlation,
respectively, for $|C| = 16$, $20$ and $24$. Specifically, EC and CBS show a
little bit of sensitivity on the evaluation of ASPL because these two
algorithms perform more Adding Edge operations than Splitting Vertex, and as
such will connect distant vertices in a large community, when the number of
detected communities is small. %
Nonetheless, the influence of the number of communities detected is
quite small for the four algorithms.

\subsubsection{Scalability}

We demonstrate the execution efficiency of our algorithms on synthetic data
sets with the number of vertices ranging from 20,000 to 100,000. The
experimental environment is a Debian GNU/Linux server with double dual-core
2.4 GHz Opteron processors and 4GB RAM. Although Figure \ref{SCL} shows that
the execution time grows as the value of $k$ increases, the proposed
algorithms can anonymize the graph to satisfy $k$-structural diversity in a
linear-time scale of the graph size.

\section{Conclusion}

In this paper, we addressed a new privacy issue, community identification,
and formulated the $k$-Structural Diversity Anonymization ($k$-SDA) problem
to protect the community identity of each individual in published social
networks. For $k$-SDA, we proposed an Integer Programming formulation to
find optimal solutions, and also devised scalable heuristics. The
experiments on real data sets demonstrated that our approaches can ensure
the $k$-structural diversity and preserve much of the characteristics of the
original social networks.

\section*{Acknowledgements}
This work is supported in part by the National Science Council of
Taiwan under Contract NSC101-2628-E-001-003-MY3 and
NSC100-2221-E-001-006-MY2, US NSF through grants IIS-0905215,
CNS-1115234, IIS-0914934, DBI-0960443, and OISE-1129076, US
Department of Army through grant W911NF-12-1-0066, Google Mobile
2014 Program, Huawei and KAU grants.



%





\ifCLASSOPTIONcaptionsoff
\newpage \fi

\end{document}